\documentclass[a4paper,10pt,onecolumn,accepted=2026-09-22]{quantumarticle}
\pdfoutput=1

\usepackage[utf8]{inputenc}  
\usepackage[english]{babel} 
\usepackage[english=american]{csquotes}
\MakeOuterQuote{"}

\usepackage{amssymb,amsthm,mathtools} 
\usepackage{hyperref} 
\usepackage[only,boxslash,llparenthesis,rrparenthesis]{stmaryrd} 
\usepackage{array}
\usepackage{physics}
\usepackage{enumitem}

\usepackage[numbers]{natbib}

\usepackage{graphicx,subcaption}  
\graphicspath{{images/}}
\usepackage[export]{adjustbox}

\usepackage{color}

\newcommand{\be}{\begin{equation}}
\newcommand{\ee}{\end{equation}}
\newcommand{\bee}{\begin{equation*}}
\newcommand{\eee}{\end{equation*}}
\newcommand{\bea}{\begin{eqnarray}}
\newcommand{\eea}{\end{eqnarray}}
\newcommand{\bean}{\begin{eqnarray*}}
\newcommand{\eean}{\end{eqnarray*}}

\newcommand{\bsub}{\begin{subequations}}
\newcommand{\esub}{\end{subequations}}
\def\bal#1\eal{\begin{align}#1\end{align}}
\def\baal#1\eaal{\begin{align*}#1\end{align*}}

\newcommand{\im}{\mathrm{i}}

\newcommand{\R}{\mathbb{R}}
\newcommand{\C}{\mathbb{C}}
\newcommand{\defeq}{\coloneqq}
\newcommand{\tens}{\otimes}
\DeclareMathOperator{\id}{id}
\newcommand{\xd}{\mathrm{d}}
\newcommand{\xD}{\mathcal{D}}
\newcommand{\cH}{\mathcal{H}}
\newcommand{\ncoh}{\mathsf{k}}
\newcommand{\mcoh}{\Xi}

\newcommand{\Lin}{\mathrm{in}}
\newcommand{\Lout}{\mathrm{out}}
\newcommand{\Li}{\mathrm{i}}
\newcommand{\Lf}{\mathrm{f}}
\newcommand{\Lint}{\mathrm{int}}
\newcommand{\Lext}{\mathrm{ext}}
\newcommand{\no}[1]{{:}#1{:}}

\newcommand{\np}{\boxslash}
\newcommand{\cB}{\mathcal{B}}
\newcommand{\discard}{\vcenter{\hbox{\includegraphics[width=1em]{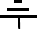}}}}
\newcommand{\cP}{\mathcal{P}}
\newcommand{\comp}{\diamond}
\newcommand{\Lren}{\mathrm{ren}}
\newcommand{\wuu}{w_{\uparrow\uparrow}}
\newcommand{\wdd}{w_{\downarrow\downarrow}}
\newcommand{\wud}{w_{\uparrow\downarrow}}
\newcommand{\wdu}{w_{\downarrow\uparrow}}
\newcommand{\wtuu}{%
  \mathop{\mathtt{w}_{\uparrow\uparrow}}\nolimits
}
\newcommand{\wtdd}{%
  \mathop{\mathtt{w}_{\downarrow\downarrow}}\nolimits
}
\newcommand{\wtud}{\mathtt{w}_{\uparrow\downarrow}}

\newcommand{\rp}{\mathtt{r}}
\newcommand{\wD}{\widetilde{D}}
\newcommand{\vac}{\mathbf{0}}
\newcommand{\Lini}{\mathrm{ini}}
\newcommand{\Lfin}{\mathrm{fin}}
\newcommand{\lb}{\llparenthesis}
\newcommand{\rb}{\rrparenthesis}

\allowdisplaybreaks

\begin{document}


\title{Towards local and compositional measurements in quantum field theory}
\author[1,2]{Robert Oeckl}
\email{robert@matmor.unam.mx}
\orcid{0000-0003-3447-7897}
\author[3]{Adamantia Zampeli}
\email{adamantia.zampeli@ug.edu.pl}
\orcid{0000-0002-1517-6127}

\affil[1]{Institute for Quantum Optics and Quantum Information,
Boltzmanngasse 3, 1090 Vienna, Austria}
\affil[2]{Centro de Ciencias Matemáticas,
Universidad Nacional Autónoma de México,
C.P.~58190, Morelia, Michoacán, Mexico}
\affil[3]{International Center for Theory of Quantum Technologies, University of Gdańsk, Wita Stwosza 63, 80-308 Gdańsk, Poland}

\maketitle


\begin{abstract}
A universal framework for the joint measurement of multiple localized observables in quantum field theory satisfying spacetime locality and compositionality is still lacking. We present an approach to the problem that is based on the one hand on the positive formalism, an axiomatic framework, where it is clear from the outset that we satisfy locality and compositionality, while also having a consistent probabilistic interpretation. On the other hand, the approach is based on standard tools from quantum field theory, in particular the path integral and the Schwinger-Keldysh formalism. After an overview of the conceptual foundations we introduce the modulus-square construction as a formalization of the measurement process for the expectation value of an important class of observables including quadratic observables. We show that this construction has many of the desired properties, including positivity, locality, single measurement recovery and compositionality. We introduce a renormalization scheme for the measurement of quadratic observables that also satisfies compositionality, in contrast to previous renormalization schemes. We discuss relativistic causality, confirming that measurements in our scheme are indeed localized in the spacetime regions where the underlying observables have support.
\end{abstract}



\section{Introduction}

Quantum field theory (QFT) is our most successful framework for describing matter and its interactions at the smallest scales. We can precisely predict outcomes of single asymptotic measurements of particle interactions through the powerful S-matrix. However, when it comes to joint measurements localized and distributed at different spacetime regions, a fully satisfactory framework applicable to QFT has been missing. This is to be contrasted with the situation in non-relativistic quantum mechanics for which a complete characterization of the notion of a measurement was achieved around 1970, in terms of trace-preserving, completely-positive maps on the space of self-adjoint operators on the Hilbert spaces of states \cite{Kra:statechanges}. In modern parlance, this is the basis of the notion of a \emph{quantum operation}, forming the heart of Quantum Information Theory. It goes almost without saying that this framework satisfies \emph{temporal locality}, i.e., a quantum operation only depends on the immediately preceding state and directly only influences the immediately subsequent state. Moreover, this framework is \emph{compositional} in the sense that there is a notion of joint measurement and that, furthermore, the implementation of a joint measurement is the composition of the implementations of the component measurements in the order corresponding to their temporal application (as maps). It is clear that the desiderata for a framework for measurements in the relativistic setting of QFT are substantially similar, but even more demanding. Namely, instead of mere temporal locality and causality we require full relativistic space-time locality and causality. Also, instead of compositionality only in time we require compositionality in spacetime.

Even the implementation of the non-relativistic measurement framework in QFT has not been straightforward. As famously shown by Sorkin \cite{Sor:impossible}, a large class of projection-valued measurements lead to superluminal signaling and are thus unphysical. However, following the non-relativistic path of constructing the quantum operation of a measurement by spectral decomposition from a self-adjoint operator, would precisely lead to projection valued measurements. What is more, typical point-localized field operators are highly singular, and even their smeared versions are unbounded and have continuous spectrum, making them difficult to deal with. It was shown only very recently, how quantum operations measuring field operators can be implemented avoiding superluminal signaling \cite{Oe:spectral}.\footnote{For a class of Gaussian measurements in scalar field theory it was previously shown by Jubb that these do not lead to superluminal signaling \cite{Jub:causalupdates}.} On the conceptual side this requires that the measurement extract the full continuous outcome value rather than a binned value, i.e., one where the real line is cut into bins, with corresponding projectors applied for the bins. On the technical side, the operations are implemented as limits of non-projective positive-operator-valued measures. However, this manner of describing measurement is still limited in that composition arises from the composition of operators and has the meaning of composition in time rather than in spacetime.

In part due to the historical difficulties with the measurement of observables, there is a considerable literature in QFT on approaches to local measurement centered on a modeling of the measurement apparatus. For an excellent recent review focused on causal measurement see \cite{PaFr:eliminatingimpossible}.
These approaches can be seen as developments of von Neumann's original description of measurement through an interaction of the system with an ancilla (representing the apparatus) \cite{vne:mathgrundquant}. The measurement proper (i.e.\ extraction of a definite outcome) then happens on the ancilla at a later time when it can be considered isolated from the measured system. Questions of locality and causality can then be reduced to the interaction between system and ancilla, as long as the measurement proper on the latter is deferred far enough into the future. These approaches can roughly be divided into two classes: Either the measuring ancilla is modeled as a non-relativistic system (with a usually finite-dimensional Hilbert space) or it is a field theoretic system itself.
In the first case, the prototypical model is the Unruh-deWitt detector \cite{BiDa:qftcurved} with two states of different energy, following a fixed trajectory in spacetime, coupled to a scalar field system. There is a large literature on developing this idea in many directions, adding spin and other degrees of freedom, with different types of coupling, trajectories, smearing the detector in space etc. For the present purposes most relevant is recent work that focuses on taking this serious as the basis for a local measurement theory of QFT \cite{PGGaMM:detectormeasurementqft}.
In the second case, the first formalization of the measurement process through a local interaction of the system field with another relativistic apparatus field was provided by Hellwig and Kraus \cite{HeKr:opmeasureii}. In the context of algebraic quantum field theory this line of research has been much further developed, culminating recently in a coherent framework for local measurement \cite{FeVe:qftlocalmeasure}. Both of these frameworks include notions of "update rules" that can be used to describe compositions of measurements. However, these notions of composition are in essence the time-ordered ones inherited from non-relativistic quantum mechanics. In particular, any two measurements can only be composed when the underlying spacetime regions are either purely spacelike or purely timelike related.
A further recent approach to a measurement theory of QFT is grounded in a histories type framework \cite{AHS:qftqi}. It makes connections with the detector approaches, even though in principle this approach is not detector-based. Because the Schwinger-Keldysh formalism is used, there are interesting connections to our approach that we discuss in the final Section~\ref{sec:discussion}.

We consider an approach to measurement in QFT based on the local version of the \emph{positive formalism} (PF) \cite{Oe:dmf,Oe:posfound}. This axiomatic framework allows for the local description of physics in spacetime regions. It is compositional in bringing into correspondence the composition of localized physical processes to the composition of the underlying spacetime regions. Central is the operational concept of \emph{probe} to describe processes of measurement, observation and intervention. This generalizes the temporally compositional concept of a quantum operation to a spacetime compositional setting. Crucially, this notion of composition is not limited by a requirement of causal orderability between measurement regions and does not require an "update rule" (even though such rules may be derived when appropriate). Rather, probes can be composed whenever their underlying spacetime regions are disjoint. Along with this comes a consistent generalization of the probability rules of the standard formulation of quantum theory. This clarifies in principle what the mathematical objects are that encode local measurements in QFT, how they are composed and how probabilities of outcomes are computed from them. However, this leaves open the crucial question how these objects are to be constructed to implement specific measurements. This may be seen as a problem of \emph{quantization}. For spacetime regions where no measurement takes place it has long been clear that the path integral provides an adequate quantization prescription.\footnote{Path integrals play a central role in \emph{topological quantum field theory} \cite{Ati:tqft} and the closely related amplitude formalism, also known as \emph{general boundary quantum field theory} \cite{Oe:gbqft}. The latter is the pure state counterpart of the local positive formalism.} In contrast, it has been an open question how to construct probes encoding the measurement of specific classical observables in a general spacetime region.

We propose a scheme for constructing probes to measure those observables which can be represented as the \emph{modulus-square} of other simpler observables. This includes correlation functions such as $\phi^2(x)$ and their products, but also for example the energy-momentum tensor. Our scheme utilizes a double path integral and in this way takes advantage of its known locality properties. In the case that the system is discarded after a joint measurement\footnote{In the context of the modulus-square construction "joint measurement" is to be understood in the sense of "joint expectation", but not "joint probability".}, this amounts to the \emph{Schwinger-Keldysh formalism}, which has long been recognized to provide a sensible setting for describing measurements in QFT of open systems \cite{CalHu:noneqqft}. We demonstrate that our proposal satisfies a number of desiderata, including positivity (which ensures coherence with the probability interpretation), locality and compositionality in spacetime. Also, as expected, the results for single measurements, which can be dealt with already in the standard QFT formalism are recovered.
For probes encoding quadratic observables, which are often singular, we introduce a \emph{renormalization} prescription. We show, that this prescription makes results not only finite, but is \emph{semiclassical}. That is, it recovers classical expectation values on coherent states. What is more striking, however, is that our renormalization prescription is \emph{compositional}, in contrast to standard prescriptions such as normal ordering. That is, renormalization "commutes" with composition, as we show. 

We demonstrate the modulus-square construction in Klein-Gordon theory in a globally hyperbolic spacetime for quadratic field observables, showing how the renormalization prescription can be justified via point-splitting regularization not only at the individual, but also at the composite level. We extend the application to the energy-momentum tensor and its simplest correlation function. By construction, all expectation values are real, and we show explicitly that they satisfy relativistic causality.

This paper is organized as follows: In Section~\ref{sec:local_measurements}, we give a brief overview of the description of the compositional frameworks of measurement in the standard formulation of quantum theory (SFQ) on the one hand, and in the positive formalism (PF) on the other hand. Section~\ref{sec:probes_qft} lays out the central notions of probe and composition in quantum field theory (QFT). The well-understood case of single measurements is briefly reviewed in Section~\ref{sec:single_meas}. We lay out our proposal of the modulus-square quantization prescription for probes in QFT in Section~\ref{sec:modsquare}, including motivation, properties and renormalization (the latter for quadratic observables). While the treatment up to this point is for general bosonic field theories we specialize in Section~\ref{sec:scalar} to scalar field theory. This provides the opportunity to consider explicit examples of observables and composition including field operators and the energy-momentum tensor. Moreover, it permits a concrete demonstration of relativistic causality. A discussion and outlook section (Section~\ref{sec:discussion}) concludes the paper, except for appendices on propagators for field observables (Appendix~\ref{sec:vacobs}) and supplementary calculations with modulus-square probes (Appendix~\ref{sec:derivations}).


\section{Local measurements and composition in the positive formalism}
\label{sec:local_measurements}

In the present section we review the positive formalism (PF) in its most basic version as a framework to encode measurements and more general operations in quantum theory (PF+Q), with an emphasis on composition. To make this treatment more accessible and emphasize the key differences, we first review relevant aspects of measurements and operations and their compositions in the standard formulation of quantum theory (SFQ). To this end we employ a diagrammatic language that is widely used in quantum foundations, categorical quantum mechanics and quantum information theory.

\subsection{Composition in the SFQ}
\label{sec:sfq}

\begin{figure}[th]
    \centering
    \begin{tabular}{m{0.2\textwidth}m{0.73\textwidth}}
\begingroup%
  \makeatletter%
  \providecommand\color[2][]{%
    \errmessage{(Inkscape) Color is used for the text in Inkscape, but the package 'color.sty' is not loaded}%
    \renewcommand\color[2][]{}%
  }%
  \providecommand\transparent[1]{%
    \errmessage{(Inkscape) Transparency is used (non-zero) for the text in Inkscape, but the package 'transparent.sty' is not loaded}%
    \renewcommand\transparent[1]{}%
  }%
  \providecommand\rotatebox[2]{#2}%
  \newcommand*\fsize{\dimexpr\f@size pt\relax}%
  \newcommand*\lineheight[1]{\fontsize{\fsize}{#1\fsize}\selectfont}%
  \ifx\svgwidth\undefined%
    \setlength{\unitlength}{63.84926208bp}%
    \ifx\svgscale\undefined%
      \relax%
    \else%
      \setlength{\unitlength}{\unitlength * \real{\svgscale}}%
    \fi%
  \else%
    \setlength{\unitlength}{\svgwidth}%
  \fi%
  \global\let\svgwidth\undefined%
  \global\let\svgscale\undefined%
  \makeatother%
  \begin{picture}(1,1.9004985)%
    \lineheight{1}%
    \setlength\tabcolsep{0pt}%
    \put(0,0){\includegraphics[width=\unitlength,page=1]{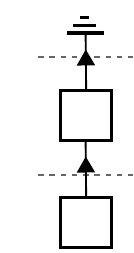}}%
    \put(0.4776154,0.99198514){\color[rgb]{0,0,0}\makebox(0,0)[lt]{\lineheight{0}\smash{\begin{tabular}[t]{l}$Q[j]$\end{tabular}}}}%
    \put(0.4776154,0.19843302){\color[rgb]{0,0,0}\makebox(0,0)[lt]{\lineheight{0}\smash{\begin{tabular}[t]{l}$\;\;\, \sigma$\end{tabular}}}}%
    \put(0,0){\includegraphics[width=\unitlength,page=2]{sqt_diagts_simple.pdf}}%
    \put(-0.00336486,1.80530359){\color[rgb]{0,0,0}\makebox(0,0)[lt]{\lineheight{1.25}\smash{\begin{tabular}[t]{l}time\end{tabular}}}}%
  \end{picture}%
\endgroup%
 &
    \begin{tabular}{m{0.15\textwidth}m{0.15\textwidth}m{0.15\textwidth}m{0.15\textwidth}}
\begingroup%
  \makeatletter%
  \providecommand\color[2][]{%
    \errmessage{(Inkscape) Color is used for the text in Inkscape, but the package 'color.sty' is not loaded}%
    \renewcommand\color[2][]{}%
  }%
  \providecommand\transparent[1]{%
    \errmessage{(Inkscape) Transparency is used (non-zero) for the text in Inkscape, but the package 'transparent.sty' is not loaded}%
    \renewcommand\transparent[1]{}%
  }%
  \providecommand\rotatebox[2]{#2}%
  \newcommand*\fsize{\dimexpr\f@size pt\relax}%
  \newcommand*\lineheight[1]{\fontsize{\fsize}{#1\fsize}\selectfont}%
  \ifx\svgwidth\undefined%
    \setlength{\unitlength}{26.03728731bp}%
    \ifx\svgscale\undefined%
      \relax%
    \else%
      \setlength{\unitlength}{\unitlength * \real{\svgscale}}%
    \fi%
  \else%
    \setlength{\unitlength}{\svgwidth}%
  \fi%
  \global\let\svgwidth\undefined%
  \global\let\svgscale\undefined%
  \makeatother%
  \begin{picture}(1,2.10198695)%
    \lineheight{1}%
    \setlength\tabcolsep{0pt}%
    \put(0,0){\includegraphics[width=\unitlength,page=1]{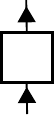}}%
    \put(0.3230662,0.94061053){\color[rgb]{0,0,0}\makebox(0,0)[lt]{\lineheight{0}\smash{\begin{tabular}[t]{l}$A$\end{tabular}}}}%
  \end{picture}%
\endgroup%
 &
\begingroup%
  \makeatletter%
  \providecommand\color[2][]{%
    \errmessage{(Inkscape) Color is used for the text in Inkscape, but the package 'color.sty' is not loaded}%
    \renewcommand\color[2][]{}%
  }%
  \providecommand\transparent[1]{%
    \errmessage{(Inkscape) Transparency is used (non-zero) for the text in Inkscape, but the package 'transparent.sty' is not loaded}%
    \renewcommand\transparent[1]{}%
  }%
  \providecommand\rotatebox[2]{#2}%
  \newcommand*\fsize{\dimexpr\f@size pt\relax}%
  \newcommand*\lineheight[1]{\fontsize{\fsize}{#1\fsize}\selectfont}%
  \ifx\svgwidth\undefined%
    \setlength{\unitlength}{26.03728731bp}%
    \ifx\svgscale\undefined%
      \relax%
    \else%
      \setlength{\unitlength}{\unitlength * \real{\svgscale}}%
    \fi%
  \else%
    \setlength{\unitlength}{\svgwidth}%
  \fi%
  \global\let\svgwidth\undefined%
  \global\let\svgscale\undefined%
  \makeatother%
  \begin{picture}(1,1.55167701)%
    \lineheight{1}%
    \setlength\tabcolsep{0pt}%
    \put(0,0){\includegraphics[width=\unitlength,page=1]{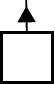}}%
    \put(0.3230662,0.39030059){\color[rgb]{0,0,0}\makebox(0,0)[lt]{\lineheight{0}\smash{\begin{tabular}[t]{l}$B$\end{tabular}}}}%
  \end{picture}%
\endgroup%
 &
\begingroup%
  \makeatletter%
  \providecommand\color[2][]{%
    \errmessage{(Inkscape) Color is used for the text in Inkscape, but the package 'color.sty' is not loaded}%
    \renewcommand\color[2][]{}%
  }%
  \providecommand\transparent[1]{%
    \errmessage{(Inkscape) Transparency is used (non-zero) for the text in Inkscape, but the package 'transparent.sty' is not loaded}%
    \renewcommand\transparent[1]{}%
  }%
  \providecommand\rotatebox[2]{#2}%
  \newcommand*\fsize{\dimexpr\f@size pt\relax}%
  \newcommand*\lineheight[1]{\fontsize{\fsize}{#1\fsize}\selectfont}%
  \ifx\svgwidth\undefined%
    \setlength{\unitlength}{26.03728731bp}%
    \ifx\svgscale\undefined%
      \relax%
    \else%
      \setlength{\unitlength}{\unitlength * \real{\svgscale}}%
    \fi%
  \else%
    \setlength{\unitlength}{\svgwidth}%
  \fi%
  \global\let\svgwidth\undefined%
  \global\let\svgscale\undefined%
  \makeatother%
  \begin{picture}(1,1.52386653)%
    \lineheight{1}%
    \setlength\tabcolsep{0pt}%
    \put(0,0){\includegraphics[width=\unitlength,page=1]{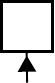}}%
    \put(0.3230662,0.94061053){\color[rgb]{0,0,0}\makebox(0,0)[lt]{\lineheight{0}\smash{\begin{tabular}[t]{l}$C$\end{tabular}}}}%
  \end{picture}%
\endgroup%
 &
        \includegraphics{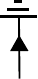} \\
        $A:\cB\to\cB$ & $B:\R\to\cB$ \newline $B\in\cB$ & $C:\cB\to\R$ & $\discard:\sigma\mapsto \tr(\sigma)$ \\
        operation & preparation & effect & discard
    \end{tabular}
\end{tabular}
    \caption{Basic elements and compositionality in the SFQ. In the case of a preparation note that the choice of a linear map $B:\R\to\cB$ is equivalent to choosing an element $B\defeq B(1)\in\cB$.}
    \label{fig:QITcomp}
\end{figure}

A quantum system is characterized primarily by its state space. Mathematically, the states live in the real vector space $\cB$ of self-adjoint operators on the Hilbert space $\cH$ of the system. Proper (but unnormalized) states are positive operators, i.e., elements of the positive cone $\cB^+\subseteq\cB$. Operations are real linear maps $A: \cB_1\to\cB_2$ between state spaces. An important property of the operations $A$ on a system is that they are required to be \emph{completely positive} maps. This property means that not only do they map the positive cone $\cB^+_1$ to the positive cone $\cB^+_2$, but it guarantees that if this system is to be seen as subsystem of a larger one, then the operation $\id \tens A$ on the composite system must also be positive. Maps with the property of complete positivity admit a decomposition in terms of Kraus operators. That is, there is a set of operators $\{K_i\}$ on $\cH$, so that,\footnote{As usual, we require $A$ to map trace-class operators to trace-class operators. This implies that $A$ admits a decomposition as indicated, see e.g.\ \cite{Hol:statquant}.}
\be\label{eq:kraus_decomp}
A(\sigma) = \sum_i K_i \sigma K^\dag_i .
\ee
Conversely, a map constructed from Kraus operators is completely positive.

Measurements are encoded by operations. If no measured value is recorded, the corresponding operation is called \emph{non-selective}. For each possible outcome of the measurement there is also a \emph{selective} operation. Crucially, the sum (as linear maps) of the selective operations over all possible outcomes yields the non-selective operation.

The operations can be represented as input-output boxes, see Figure~\ref{fig:QITcomp}. Particular operations are the \emph{preparations} that take no input and give an output and the \emph{effects} that accept an input and produce a scalar output, i.e., a complex number. That is, the effects can be viewed as positive elements of the vector space $\cB^*$, that is dual to the corresponding state space.
Moreover, there is a distinguished element on $\cB^*$ which in the following we call the \emph{discard (effect)}. This is defined as $\discard: \cB \rightarrow \R, \ \discard(\cdot) \equiv \tr (\cdot)$; the states $\sigma\in\cB^+$ for which $\discard(\sigma) =1$ are called \emph{normalized}.

\begin{figure}[h]
\centering
\begin{subfigure}[t]{0.3\textwidth}
    \centering
    \includegraphics[scale=1.75,valign=t]{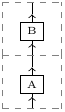}
    \subcaption{map composition}
\end{subfigure}
\centering
    \begin{subfigure}[t]{0.3\textwidth}
    \centering
\includegraphics[scale=2,valign=t]{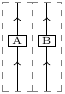}
\subcaption{tensor product}
\end{subfigure}
\begin{subfigure}[t]{0.3\textwidth}
\centering
\includegraphics[scale=2.1,valign=t]{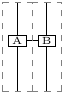}
\subcaption{generalized composition}
\end{subfigure}
\caption{Composition in horizontal and vertical directions.}
\label{fig:compositions}
\end{figure}

An important feature of the formalism is the composability of all of these elements. There are two types of composition: one is in the temporal/vertical direction depicted in the first diagram of Figure~\ref{fig:QITcomp} and the diagram of Figure~\ref{fig:compositions}.a. The composition is through the composition of the maps corresponding to the operations; while in the spatial/horizontal direction the composition is through the tensor product of the state spaces and underlying Hilbert spaces, see Figure \ref{fig:compositions}.b. 
The predictions in the SFQ are given through the probabilistic Born rule. This assumes that the system at the end is discarded. The probability for given joint measurement outcomes is then simply the real number resulting from the evaluation of the diagram composed of preparations, operations and discard effects that describe the experimental setup. Hereby the operations to be used are the selective ones encoding the various component measurement outcomes. The simplest case is a preparation of a state $\sigma$, followed by a measurement $Q$, followed by discarding the system. Say the measurement can have $n$ outcomes, and we denote the corresponding operations $Q[j]$ with $j\in\{1,\ldots,n\}$. The non-selective operation is $Q[*]=\sum_{j=1}^n Q[j]$ and the probability $\Pi$ to register outcome $j$ is given by
\begin{equation}
    \Pi=\tr(Q[j](\sigma)) .
\end{equation}
The corresponding graphical representation is the diagram on the left-hand side of Figure~\ref{fig:QITcomp}.

\begin{figure}[ht]
    \centering
    \begin{tabular}{cc}
        $\vcenter{\hbox{
\begingroup%
  \makeatletter%
  \providecommand\color[2][]{%
    \errmessage{(Inkscape) Color is used for the text in Inkscape, but the package 'color.sty' is not loaded}%
    \renewcommand\color[2][]{}%
  }%
  \providecommand\transparent[1]{%
    \errmessage{(Inkscape) Transparency is used (non-zero) for the text in Inkscape, but the package 'transparent.sty' is not loaded}%
    \renewcommand\transparent[1]{}%
  }%
  \providecommand\rotatebox[2]{#2}%
  \newcommand*\fsize{\dimexpr\f@size pt\relax}%
  \newcommand*\lineheight[1]{\fontsize{\fsize}{#1\fsize}\selectfont}%
  \ifx\svgwidth\undefined%
    \setlength{\unitlength}{26.03728731bp}%
    \ifx\svgscale\undefined%
      \relax%
    \else%
      \setlength{\unitlength}{\unitlength * \real{\svgscale}}%
    \fi%
  \else%
    \setlength{\unitlength}{\svgwidth}%
  \fi%
  \global\let\svgwidth\undefined%
  \global\let\svgscale\undefined%
  \makeatother%
  \begin{picture}(1,2.92255287)%
    \lineheight{1}%
    \setlength\tabcolsep{0pt}%
    \put(0,0){\includegraphics[width=\unitlength,page=1]{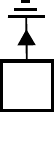}}%
    \put(0.05397673,1.19172802){\color[rgb]{0,0,0}\makebox(0,0)[lt]{\lineheight{0}\smash{\begin{tabular}[t]{l}$\;\; A$\end{tabular}}}}%
    \put(0,0){\includegraphics[width=\unitlength,page=2]{sqt_diag_op_e.pdf}}%
  \end{picture}%
\endgroup%
}}\;\;=\; \vcenter{\hbox{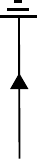}}$ &
        $\vcenter{\hbox{
\begingroup%
  \makeatletter%
  \providecommand\color[2][]{%
    \errmessage{(Inkscape) Color is used for the text in Inkscape, but the package 'color.sty' is not loaded}%
    \renewcommand\color[2][]{}%
  }%
  \providecommand\transparent[1]{%
    \errmessage{(Inkscape) Transparency is used (non-zero) for the text in Inkscape, but the package 'transparent.sty' is not loaded}%
    \renewcommand\transparent[1]{}%
  }%
  \providecommand\rotatebox[2]{#2}%
  \newcommand*\fsize{\dimexpr\f@size pt\relax}%
  \newcommand*\lineheight[1]{\fontsize{\fsize}{#1\fsize}\selectfont}%
  \ifx\svgwidth\undefined%
    \setlength{\unitlength}{71.13158275bp}%
    \ifx\svgscale\undefined%
      \relax%
    \else%
      \setlength{\unitlength}{\unitlength * \real{\svgscale}}%
    \fi%
  \else%
    \setlength{\unitlength}{\svgwidth}%
  \fi%
  \global\let\svgwidth\undefined%
  \global\let\svgscale\undefined%
  \makeatother%
  \begin{picture}(1,1.06978287)%
    \lineheight{1}%
    \setlength\tabcolsep{0pt}%
    \put(0,0){\includegraphics[width=\unitlength,page=1]{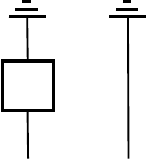}}%
    \put(0.51611233,0.43659906){\color[rgb]{0,0,0}\makebox(0,0)[lt]{\lineheight{1.25}\smash{\begin{tabular}[t]{l}$=$\end{tabular}}}}%
    \put(0,0){\includegraphics[width=\unitlength,page=2]{sqt_diagn_opeqe_crossed.pdf}}%
    \put(0.10532013,0.43622486){\color[rgb]{0,0,0}\makebox(0,0)[lt]{\lineheight{0}\smash{\begin{tabular}[t]{l}$A$\end{tabular}}}}%
  \end{picture}%
\endgroup%
}}$ \\ \\
        $\tr(A(\sigma)) = \tr(\sigma)$ &
        \phantom{$\tr(A(\sigma)) \neq \tr(\sigma)$} \\ \\
        $A:\cB\to\cB$ \emph{non-selective} &
        $A:\cB\tens\cB\to\R$ \emph{non-selective}
    \end{tabular}
    \caption{Causality axiom.}
    \label{fig:causality_axiom}
\end{figure}

Ensuring consistency of the probability rule requires the \emph{causality axiom}. Mathematically, this stipulates that non-selective operations, i.e., operations where no measured value is extracted, are \emph{trace-preserving}. That is, they should satisfy the equality $\discard(A(\sigma))=\discard(\sigma)$ for any state $\sigma\in\cB$. This is depicted in Figure~\ref{fig:causality_axiom}, left-hand side. In terms of the Kraus operators determining $A$, see \eqref{eq:kraus_decomp}, this condition reads 
\be\label{trace-pres}
\sum_i K^\dag_i K_i = \id .
\ee 
Considering preparations as a special type of operation where the final state space is $\R$, this condition translates to $\discard(\sigma)=1$. This is the already mentioned normalization condition for states.
Imposing the trace-preservation condition on non-selective operations enforces
an important and reasonable constraint in our physical theories that states that no future choice (such as choosing what is to be measured) must influence the present. In this form, it states that performing a non-selective measurement and then discarding the system should give the same probability/predictions as just discarding the system without performing any measurement. 

In the case of a finite dimensional Hilbert space $\cH$, the space $\cB$ is identified with the space of self-adjoint operators on $\cH$. The dual space $\cB^*$ can be identified with the space $\cB$ through the Hilbert-Schmidt inner product. That is, we can identify $\tau\in\cB$ with an element in $\cB^*$ by setting,
\begin{equation}
    \tau(\sigma)=\tr(\tau^\dagger \sigma) .
    \label{eq:pairing}
\end{equation}
In the infinite-dimensional case we take advantage of the normalization condition on states and restrict $\cB$ to be the space of self-adjoint operators that are \emph{trace-class}. On the other hand we identify elements of $\cB^*$ with self-adjoint operators that are merely bounded, by the same pairing \eqref{eq:pairing}. This is necessary as for example the discard effect encoded by the trace corresponds in this way to the identity operator, which is not trace-class in infinite dimensions. Note that this introduces an asymmetry between the spaces $\cB$ and $\cB^*$.

\subsection{The positive formalism}
\label{sec:posform}

The SFQ presupposes a fixed notion of time. Initial and final state spaces are distinguished as domains and images of the maps encoding operations. In the diagrams this is indicated by the arrows on the links, see Figures~\ref{fig:QITcomp}, \ref{fig:compositions}.a and \ref{fig:compositions}.b. This is also reflected in the correspondence of the different types of composition to different temporal relations. Horizontal composition encodes a simultaneous combination of different systems and operations, while vertical composition encodes the consecutive, i.e., temporally sequential combination of operations on systems.

In relativistic physics (such as in QFT), the temporal order of events may be frame dependent and a formalism that depends on such an order is cumbersome at best and inapplicable at worst. To address this, we shall use the \emph{positive formalism} (PF) instead of the SFQ. More precisely, we shall use the quantum theory variant of the PF, also referred to as PF+Q. The PF in general is a framework that applies not only to quantum theory, but also to classical theory and even generalized probabilistic theories \cite{Oe:dmf,Oe:posfound}. Moreover, there are different versions of the PF depending on the underlying notion of spacetime (if any), see below. The version we denote as PF and thus also its quantum variant PF+Q does not presuppose any notion of space or time, but merely an abstract notion of composability.

Taking as a starting point the SFQ, the transition to the PF+Q is easy to explain, at an intuitive level: Remove the arrows from the links. Physically speaking, links still represent "systems" or, more abstractly, a means of signaling or interaction between operations, but not a priori with a specific time direction. Basic operations may still be represented as boxes, except that there is no longer a specific distinction between incoming and outgoing links. 
In order to accommodate this and other generalizations we introduce the notion of \emph{probe}.
Mathematically speaking, instead of encoding an operation as a linear map from a tensor product of incoming to a tensor product of outgoing state spaces we may encode a probe as a map from a tensor product of state spaces to the real numbers (i.e., as if all state spaces were incoming), $\cP:\cB_1\tens\cdots\tens\cB_n\to\R$. 
We note that a map $\cP$ defined in this way is \emph{positive} if and only if any partially dualized version of $\cP$ (i.e., where some in-links are converted to out-links) is completely positive. In that case, we call the corresponding probe \emph{primitive}. Thus, operations in SFQ give rise to primitive probes in PF+Q. Composition is achieved by inserting complete bases of the state spaces $\cB$ associated to the links that implement the composition.\footnote{This is the composition formula for quantum operations $\lb\tau, B(A(\sigma))\rb=\sum_{k} \lb\tau, B(\xi_k)\rb \lb\xi_k, A(\sigma)\rb$, where $\{\xi_k\}$ is an orthonormal basis with respect to the Hilbert-Schmidt inner product $\lb\tau,\sigma\rb=\tr(\tau^\dagger \sigma)$.} It is evident that in this way any composition of operations in the SFQ can be obtained by composing corresponding primitive probes in PF+Q, followed by adding the corresponding arrows. But more general compositions are possible, see Figure~\ref{fig:compositions}.c.

There is one apparent mathematical difficulty, however. This is the asymmetry between a state space $\cB$ and its dual $\cB^*$ in the infinite-dimensional case. Eliminating directionality requires the identification of $\cB$ with $\cB^*$ through the pairing \eqref{eq:pairing}. Since effects are given by bounded operators while states by trace class operators in the SFQ we are forced to choose the larger class, i.e., the bounded operators. However, the Hilbert-Schmidt inner product \eqref{eq:pairing} between two bounded operators is not well-defined in infinite dimensions. We solve this problem by taking advantage of positivity. Instead of considering the Hilbert-Schmidt inner product as a pairing $\cB\times\cB\to\R$ between vector spaces, we restrict it to the positive cones $\cB^+\times\cB^+\to\R_0^+$. Its values are then restricted to be non-negative. We may then extend $\R^+_0=[0,\infty)$ to its one-point compactification $[0,\infty]$. In this way, $\cB^+\times\cB^+\to [0,\infty]$ becomes a well-defined map, even in infinite dimensions \cite{Oe:posfound}. Note that this makes sense as physical states or effects in SFQ are always positive. Moreover, operations are completely positive, making their undirected versions positive, i.e., the probes primitive. In this way, all compositions respect positivity and can be carried out to yield a well-defined result in $[0,\infty]$, for any closed diagram. But what is the interpretation if the result is $\infty$? Either the setup is unphysical in the sense of being underdetermined\footnote{This was discussed for the \emph{general boundary formulation}, a predecessor of the PF, already in \cite{Oe:gbqft}.} or renormalization is required (see below).

There is more to the SFQ than we have taken into account so far: Crucially, there is the probability rule and there is the causality axiom. Let's start with the latter, see Figure~\ref{fig:causality_axiom} left-hand side. It would be tempting to introduce in the PF+Q a corresponding axiom, by simply dropping the arrows, see Figure~\ref{fig:causality_axiom} right-hand side. However, returning to the SFQ, by reintroducing arrows, this relation would be equivalent to imposing all relations generated by all possible choices of arrows. There are just two choices, one corresponding to the original axiom (left-hand side of Figure~\ref{fig:causality_axiom}) and its time-reversed version. Now, the latter is known to be too strong and making SFQ essentially trivial \cite{CoGoSe:timerevcausalnoise}. In conclusion, introducing the axiom of the right-hand side of Figure~\ref{fig:causality_axiom} in PF+Q would be too strong and is not supported by known quantum physics. As a consequence, there is no specific mathematical condition that singles out non-selective probes and there is in general no notion of normalization for "states".

The causality axiom is required in the SFQ, however, for the consistency of the probability interpretation. So, how can we talk about probabilities in the PF+Q?
The answer is that the PF (and thus PF+Q) has a probability rule that does not rely on the directedness coming from a notion of time. In short, we need two copies of a diagram, evaluate both, and obtain the probability as the quotient of the two values. The first diagram is like the diagram in SFQ. Here, we use the selective probes corresponding to the measurement outcomes that we are asking for. Its value becomes the numerator. The second diagram is like the first diagram, but with non-selective probes replacing the selective ones. It represents the same experiment, but without the knowledge of the outcomes. Its value becomes the numerator. The quotient is the probability. The rule can be stated more generally by saying that the denominator encodes what we know or impose about the experiment, while the numerator encodes what on top of that we ask for. In fact, the denominator does not have to be composed exclusively of non-selective probes; we may condition on partial outcomes with the very same rule.

\begin{figure}[ht]
    \centering
    \includegraphics[width=0.4\textwidth]{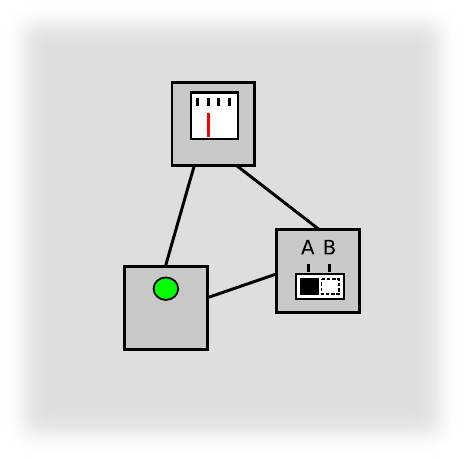}
    \includegraphics[width=0.4\textwidth]{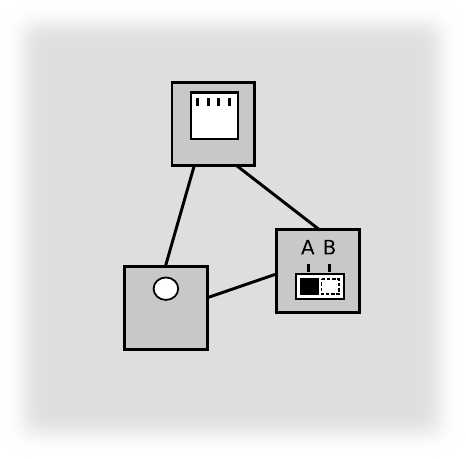}
    \caption{Probabilities are relative, arise from quotients of diagrams.}
    \label{fig:relative_probabilities}
\end{figure}

Consider the example of the situation depicted in Figure~\ref{fig:relative_probabilities}. There are three connected devices: One is a switch with two positions, A and B. One carries an indicator light that may show (g)reen or (r)ed. One is a pointer device with a scale. We are asking for the probability of a specific readout $k$ on the pointer device and for the light to be (g)reen, given that the switch is in position A. The left-hand side of Figure~\ref{fig:relative_probabilities} depicts this specific configuration. The boxes stand for primitive probes that we denote $Q[A]$ (switch), $P[g]$ (light) and $R[k]$ (pointer device). The right-hand side of Figure~\ref{fig:relative_probabilities} depicts the corresponding configuration that reflects only what we know already or have fixed about the experiment. Here, the color of the light and the pointer position are undetermined. The corresponding primitive probes are denoted $P[*]$ (light) and $R[*]$ (pointer device). These are non-selective probes, while their selective counterparts are $P[g]$ and $R[k]$, as already mentioned. The probability $\Pi$ is thus the quotient
\be\label{eq:general_prob}
\Pi = \frac{P[g] \diamond Q[A] \diamond R[k]}{P[*]\diamond Q[A]\diamond R[*]} .
\ee
Here, the numerator represents the value of the diagram on the left-hand side of Figure~\ref{fig:relative_probabilities} and the denominator the value of the diagram on the right-hand side. We use the notation $\diamond$ to generically indicate composition. Note that composition can be performed successively and is associative in this sense.

\begin{figure}[ht]
    \centering
    \includegraphics[width=\textwidth]{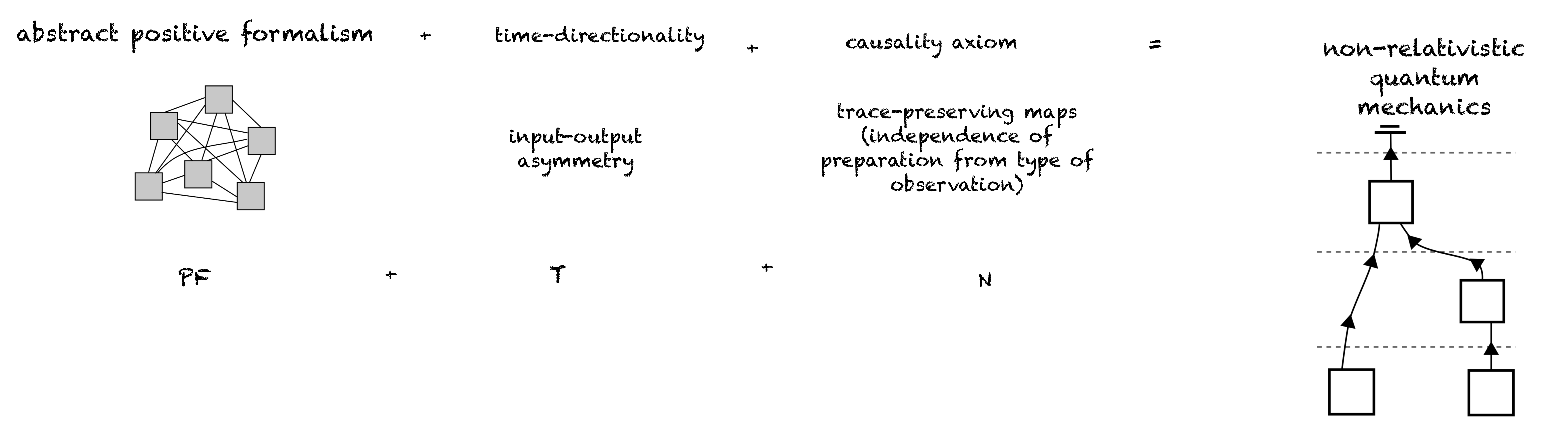}
    \caption{Recovery of SFQ from PF+Q if we re-introduce time arrows and the causality axiom.}
    \label{fig:qit_diagram}
\end{figure}

The probability rule of the PF+Q is strictly a generalization of the probability rule of the SFQ. What is more, we can recover exactly the SFQ from the PF+Q by adding structure, see Figure~\ref{fig:qit_diagram}. As a first step, we introduce arrows on all links, introduce the input-output asymmetry, and prohibit compositions that yield cycles. That is, we go from graphs to acyclic directed graphs. We indicate this addition of structure as +T (for time). Then, we impose the causality axiom (left-hand side of Figure~\ref{fig:causality_axiom}). We indicate this axiomatic enhancement by +N. We claim thus that the PF+Q+T+N recovers precisely the SFQ. To see that for the probability rule, consider a situation where the SFQ rule is applicable. That is, the system is discarded after the measurement, and we do not condition on any partial measurement outcome. But this precisely implies that the denominator of the PF+Q rule is a composition of non-selective operations with the discard. Due to the causality axiom this implies that its numerical value is precisely $1$. The probability expression reduces to the value of its numerator, which is directly equal to that of the SFQ rule.

The probes we have considered so far were exclusively primitive probes, generalizing operations in SFQ. Non-primitive probes play an important role when we are interested in (possibly joint) expectation values instead of probabilities. Suppose we have a measurement with $n$ possible outcomes $k\in\{1,\ldots,n\}$. For each outcome we have a selective probe $R[k]$ while the non-selective probe is $R[*]=\sum_{k=1}^n R[k]$. All these probes are primitive. We now assign real values $\lambda_k$ to the outcomes and define the probe $R[\Lambda]=\sum_{k=1}^n \lambda_k R[k]$. This probe is not necessarily primitive as the values $\lambda_k$ need not be positive. Using the previous example (compare expression \eqref{eq:general_prob}), we may ask for example what is the expected value $\langle\Lambda\rangle$ shown on the pointer device assuming the switch in position A and the light showing green. This is,
\be
\langle\Lambda\rangle = \frac{P[g] \diamond Q[A] \diamond R[\Lambda]}{P[g]\diamond Q[A]\diamond R[*]} .
\ee


\section{Construction of probes in quantum field theory}
\label{sec:probes_qft}

\begin{figure}
    \centering
\begingroup%
  \makeatletter%
  \providecommand\color[2][]{%
    \errmessage{(Inkscape) Color is used for the text in Inkscape, but the package 'color.sty' is not loaded}%
    \renewcommand\color[2][]{}%
  }%
  \providecommand\transparent[1]{%
    \errmessage{(Inkscape) Transparency is used (non-zero) for the text in Inkscape, but the package 'transparent.sty' is not loaded}%
    \renewcommand\transparent[1]{}%
  }%
  \providecommand\rotatebox[2]{#2}%
  \newcommand*\fsize{\dimexpr\f@size pt\relax}%
  \newcommand*\lineheight[1]{\fontsize{\fsize}{#1\fsize}\selectfont}%
  \ifx\svgwidth\undefined%
    \setlength{\unitlength}{161.47818375bp}%
    \ifx\svgscale\undefined%
      \relax%
    \else%
      \setlength{\unitlength}{\unitlength * \real{\svgscale}}%
    \fi%
  \else%
    \setlength{\unitlength}{\svgwidth}%
  \fi%
  \global\let\svgwidth\undefined%
  \global\let\svgscale\undefined%
  \makeatother%
  \begin{picture}(1,1.12633993)%
    \lineheight{1}%
    \setlength\tabcolsep{0pt}%
    \put(0,0){\includegraphics[width=\unitlength,page=1]{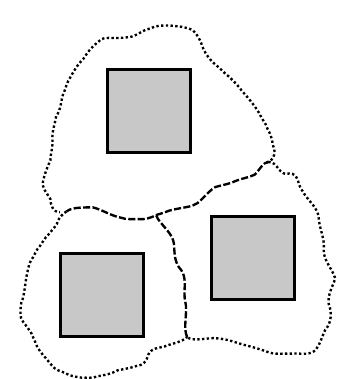}}%
    \put(0.37385952,0.54445887){\color[rgb]{0,0,0}\makebox(0,0)[lt]{\lineheight{0}\smash{\begin{tabular}[t]{l}$\Sigma_{1 3}$\end{tabular}}}}%
    \put(0.598513,0.63463173){\color[rgb]{0,0,0}\makebox(0,0)[lt]{\lineheight{0}\smash{\begin{tabular}[t]{l}$\Sigma_{2 3}$\end{tabular}}}}%
    \put(0.44427352,0.23560443){\color[rgb]{0,0,0}\makebox(0,0)[lt]{\lineheight{0}\smash{\begin{tabular}[t]{l}$\Sigma_{1 2}$\end{tabular}}}}%
    \put(0,0){\includegraphics[width=\unitlength,page=2]{devices_linked_st.pdf}}%
    \put(0.08844178,0.18882213){\color[rgb]{0,0,0}\makebox(0,0)[lt]{\lineheight{0}\smash{\begin{tabular}[t]{l}$\Sigma_1$\end{tabular}}}}%
    \put(0.45989738,0.99283842){\color[rgb]{0,0,0}\makebox(0,0)[lt]{\lineheight{0}\smash{\begin{tabular}[t]{l}$\Sigma_3$\end{tabular}}}}%
    \put(0.77024763,0.13254094){\color[rgb]{0,0,0}\makebox(0,0)[lt]{\lineheight{0}\smash{\begin{tabular}[t]{l}$\Sigma_2$\end{tabular}}}}%
    \put(0,0){\includegraphics[width=\unitlength,page=3]{devices_linked_st.pdf}}%
  \end{picture}%
\endgroup%

    \caption{Probes are localized in spacetime regions. The links between probes which carry the state spaces, are associated to hypersurfaces.}
    \label{fig:local_pf}
\end{figure}

In order to apply the PF+Q to QFT we take advantage of an important fundamental structure of the latter: spacetime. That is, we take probes to be localized in spacetime, and concretely in spacetime \emph{regions}, see Figure~\ref{fig:local_pf}. What is more, by \emph{locality}, communication or interaction between regions should only be possible through shared boundary hypersurfaces. In particular, links are associated to hypersurfaces. Moreover, again by locality, the state space associated to a link should depend only on the underlying hypersurface and not on the probes it may connect. That is, to any (oriented) hypersurface $\Sigma$ we associate a state space $\cB_{\Sigma}$. On the other hand, the probes that "fit" in a specific spacetime region generally depend on that region. That is, to any spacetime region $M$ we associate a space of probes $\cP_M$. A priori, a probe in $M$ communicates through a single state space $\cB_{\partial M}$, where $\partial M$ is the boundary hypersurface of $M$. However, when a hypersurface $\Sigma$ (like this boundary) is subdivided into hypersurface components $\Sigma=\Sigma_1\cup\cdots\cup\Sigma_n$, the state space splits into a tensor product corresponding to the components $\cB_{\partial M}=\cB_1\tens\cdots\tens\cB_n$.\footnote{This is only the simplest composition rule, adequate for the purposes of the present article. In general this rule has to be refined for gauge theories \cite{Oe:2dqym,DiOe:qabym}, and due to vacuum entanglement \cite{CoOe:locgenvac}.} There is also a special type of primitive probe for each region $M$, called the \emph{null probe} and denoted $\np_M\in\cP_M$. This probe represents the absence of any measurement or intervention, i.e., "free evolution" in the corresponding region. The null probe has the special property that it composes to itself, i.e., if $M$ and $N$ are regions, $\np_{M\cup N}=\np_M\diamond\np_N$. In SFQ the analog of the null probe is a free time evolution between operations. The reason this does not usually merit a specific representation is that one can easily absorb this into the operations, e.g., by adopting a Heisenberg picture. This is not possible in the present setting, where every spacetime region is a priori different. The framework we have arrived at in this way \cite{Oe:posfound} is denoted PF+Q+LOC, with +LOC for \emph{locality}. For simplicity, we shall refer to it as the \emph{local} PF, since we are exclusively considering quantum theory in this work.
In the following we shall see how we can employ and generalize the tools of QFT to construct and work with the relevant objects of the local PF.

\subsection{Composition in QFT}
\label{sec:compqft}

In QFT, scattering processes are the main way to extract information through measurements of cross-sections, see e.g.\ \cite{PeSc:qft}. These are given in terms of transition amplitudes of the type $\langle \psi_{\Lf}^{\Lout} |\psi_{\Li}^{\Lin} \rangle$ and are calculated by employing the $S$-matrix picture, the structure of which can be summarized as follows: i) the spacetime (usually Minkowski space) admits a global time orientation through a selection of a spacelike foliation, ii) at the far past ($t \rightarrow - \infty$) and at the late future ($t \rightarrow + \infty$) the field $\phi (x)$ is free, iii) the interaction takes place at intermediate times, iv) the free theories at $t \rightarrow \pm \infty$ are related through the $S$-matrix, and we can write 
\be\label{trans_ampl_oper_s-matrix}
\langle \psi_{\Lf}^{\mathrm{out}} |\psi_{\Li}^{\mathrm{in}} \rangle = \langle \psi_{\Lf}^{\mathrm{in}} | S | \psi_{\Li}^{\mathrm{in}} \rangle 
= \lim_{\substack{t_1 \rightarrow -\infty \\ t_2 \rightarrow + \infty}} \langle \psi_{\Lf} | U_{[t_1,t_2]}| \psi_{\Li} \rangle .
\ee
The $S$-matrix contains all the information about the interactions. The Hilbert spaces at $t \rightarrow \pm \infty$ are related through an isomorphism and can be considered "copies" of each other. The probability is then given by 
\be\label{trans_prob}
\Pi = |\langle \psi_{\Lf} | S| \psi_{\Li} \rangle|^2.
\ee
The transition amplitude \eqref{trans_ampl_oper_s-matrix} can be recast in the path integral form. For the finite-time amplitude, this is
\be
\rho_{[t_1,t_2]} (\psi_{\Li} \otimes \psi_{\Lf})
\defeq \langle \psi_{\Lf} | U_{[t_1,t_2]}| \psi_{\Li} \rangle
=\int_{K_{[t_1,t_2]}} \xD\phi\, \psi_{\Li} (\phi|_{t_1})\overline{\psi_{\Lf} (\phi|_{t_2})} e^{iS_{[t_1,t_2]} (\phi)},
\label{eq:tampl}
\ee
where $K_{[t_1,t_2]}$ is the space of field configurations in the interior, while $S_{[t_1,t_2]} (\phi)$ is the classical action functional defined on this space of field configurations. $\phi|_t$ is the induced field configuration at time $t$ and $\psi(\phi|_t)$ is the Schrödinger wave function of the state $\psi$ evaluated on this field configuration. Note that here and in the following we write for simplicity $[t_1,t_2]$ instead of $[t_1,t_2]\times\R^3$ to denote spacetime regions that are time-intervals extended over all of space.

This notion of amplitude can be extended from time interval regions $[t_1,t_2]$ to arbitrary spacetime regions $M\subseteq \R\times\R^3$ if we rewrite \eqref{eq:tampl} in the form \cite{Oe:gbqft}
\be\label{amplit_map}
\rho_M (\psi) = \int_{K_M} \xD\phi\, \psi (\phi|_{\partial_M}) e^{i S_M(\phi)}.
\ee
The map $\rho_M : \mathcal{H}_{\partial M} \rightarrow \mathbb{C}$ is a linear map from the Hilbert space $\mathcal{H}_{\partial M}$ at the boundary $\partial M$ of $M$ to the complex numbers, and we have written $\phi|_{\partial M}$ for the field configuration on the boundary. This extension of QFT is called general boundary quantum field theory (GBQFT) \cite{Oe:gbqft}.\footnote{We caution the reader that the program of GBQFT is still incomplete with respect to textbook QFT. That is, Hilbert spaces and amplitudes for regions that are different from time-interval regions have so far been constructed only for certain classes of theories. Indeed, it is clear that in cases where hypersurfaces have corners, a modification of the original axioms of \cite{Oe:gbqft} is needed, see also \cite{Oe:locqft,CoOe:locgenvac}. However, we only consider time-interval regions in this article, so this need not concern us.} The previous case of a transition amplitude is recovered by noting that in that case the boundary $\partial M$ decomposes into two pieces, the equal-time hypersurfaces at $t_1$ and $t_2$ and consequently $\psi=\psi_{\Li}\tens\psi_{\Lf}^*$.
In general, the boundary and the corresponding Hilbert space of states does not have to consist of two components. Rather, the state space is considered unified a priori and any kind of decomposition might emerge later due to additional structure and for physical considerations (e.g.\ type of the field). It is easy to see that the amplitude satisfies a composition law due to the composability of path integrals, i.e.\ if $M$ is a disjoint union $M = M_1 \cup M_2$, then
\be
\rho_{M_1 \cup M_2}  (\psi_{\Li} \otimes \psi_{\Lf}) = \sum_{k \in I} \rho_{M_1} (\psi_{\Li} \otimes \zeta_k) \rho_{M_2} (\zeta^*_k \otimes \psi_{\Lf}) .
\label{eq:amplgluing}
\ee
Here, $\{\zeta_k\}_{k\in I}$ is an orthonormal basis of the Hilbert space $\cH_{\Sigma}$ associated to the hypersurface $\Sigma$ where $M_1$ and $M_2$ are glued together. In the special case where $M_1$ and $M_2$ are time-interval regions, this is just the usual composition of evolution operators $U_{[t_1,t_3]}=U_{[t_2,t_3]} U_{[t_1,t_2]}$ in terms of their matrix elements,
\be
\langle \psi_{\Lf}, U_{[t_1,t_3]}\psi_{\Li} \rangle = \sum_{k \in I} \langle \psi_{\Lf}, U_{[t_2,t_3]} \zeta_k\rangle\langle \zeta_k, U_{[t_1,t_2]}\psi_{\Li}\rangle .
\ee
We shall sometimes abbreviate the composition formula by the implicit notation
\be\label{amplitude_compos}
\rho_M (\psi_{\Li} \otimes \psi_{\Lf}) = \rho_{M_1} \diamond \rho_{M_2} (\psi_{\Li} \otimes \psi_{\Lf}) ,
\ee
where $\diamond$ denotes the gluing at the boundary of the two regions performed through the summation over the orthonormal basis $\{\zeta_k\}_{k\in I}$ inserted at their common boundary.

\begin{figure}[t!]
    \centering
    \begin{subfigure}[t]{0.45\textwidth}
    \centering
\includegraphics[scale=0.7]{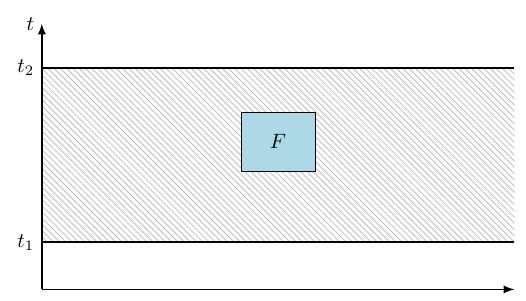}
\subcaption{An observable $F$ with support in the time-interval $[t_1,t_2]$.}
\end{subfigure}
\hfill
\begin{subfigure}[t]{0.45\textwidth}
\centering
\includegraphics[scale=0.7]{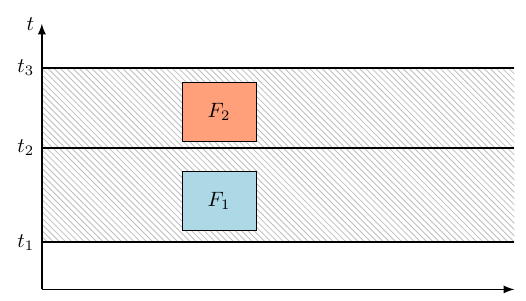}
\subcaption{Composition of two observables $F_1$ and $F_2$ in $[t_1,t_2]$ and $[t_2,t_3]$ respectively. This yields the observable $F=F_1 F_2$ in $[t_1,t_3]$.}
\end{subfigure}
\caption{Observables and their composition in time-interval regions through the path integral.}
\label{fig:obscomp}
\end{figure}

In QFT we can also insert observables into the path integral, for example to obtain the time-ordered correlation functions. Thus let $F:K_M\to\R$ be an observable defined on field configurations in the region $M$ (illustrated in Figure~\ref{fig:obscomp}.a for $M=[t_1,t_2]$). We define the correlation function
\be
\rho_M[F] (\psi) \defeq \int_{K_M} \xD\phi\, \psi (\phi|_{\partial_M})
 F(\phi) e^{\im S_M(\phi)}.
 \label{eq:piquant}
 \ee
If $M$ is a time-interval region $[t_{\Li},t_{\Lf}]$ and $F(\phi)= \phi (x_n,t_n)\cdots\phi (x_1,t_1)$, where $t_\Lf>t_k>t_\Li$, we obtain the time-ordered correlation function,
\begin{equation}
   \rho_{[t_{\Li},t_{\Lf}]}[F](\psi_{\Li}\tens\psi_{\Lf})=\bra{\psi_{\Lf}} \mathbf{T} \phi (x_n,t_n)\cdots\phi (x_1,t_1)\ket{\psi_{\Li}} .
\end{equation}
Crucially, the composition property of the path integral extends to observables inserted into it. That is, suppose we have an observable $F_1$ in region $M_1$ and an observable $F_2$ in the adjacent region $M_2$, then we obtain the joint correlation function of the product observable $F_1 F_2$ by applying the same gluing rule \eqref{eq:amplgluing},
\be
\rho_{M_1 \cup M_2}[F_1 F_2]  (\psi_{\Li} \otimes \psi_{\Lf}) = \sum_{k \in I} \rho_{M_1}[F_1] (\psi_{\Li} \otimes \zeta_k) \rho_{M_2}[F_2] (\zeta^*_k \otimes \psi_{\Lf}) .
\label{eq:obsgluing}
\ee
This is illustrated in Figure~\ref{fig:obscomp}.b.

\subsection{Composition in the local PF}
\label{sec:complocpf}

Amplitudes and correlation functions of QFT have precisely the composition properties we seek. We proceed to consider the construction of the relevant objects of the local PF from these ingredients. On the level of state spaces, the Hilbert space $\cH_{\Sigma}$ of pure states associated to a hypersurface $\Sigma$ is replaced by the corresponding space of self-adjoint operators that we shall denote by $\cB_{\Sigma}$. The positive operators form the cone $\cB_{\Sigma}^+\subseteq\cB_{\Sigma}$. As explained previously, these play the role of (unnormalized) states and effects in the SFQ.

As for probes in spacetime (generalizing quantum operations in SFQ), the first object of interest is the \emph{null probe} that corresponds to no intervention or "free evolution". It is the mixed state formalism analog of the amplitude. For a region $M$ this is the map $\np_M:\cB_{\partial M}\to\R$ given by \cite{Oe:dmf},
\begin{equation}
    \np_M(\sigma) = \sum_{k\in I} \rho_M (\sigma \zeta_k) \overline{\rho_M(\zeta_k)} .
    \label{eq:ampltonp}
\end{equation}
Again, $\{\zeta_k\}_{k\in I}$ denotes an orthonormal basis of the Hilbert space $\cH_{\partial M}$. As shown in \cite{Oe:dmf}, $\np_M$ is a primitive probe and thus positive. In particular, if $\sigma$ is positive, i.e.\ $\sigma\in\cB_{\partial M}^+$, then $\np_M(\sigma)\ge 0$.
Expression \eqref{eq:ampltonp} might look unfamiliar. However, in the case of a time-interval region, compare expression \eqref{eq:tampl}, we can rewrite it as,
\begin{equation}
    \np_{[t_1,t_2]}(\sigma_{\Li}\tens\sigma_{\Lf})
    =\tr\left(\sigma_{\Lf} U_{[t_1,t_2]} \sigma_{\Li} U_{[t_1,t_2]}^\dagger\right) .
    \label{eq:evoltonp}
\end{equation}
That is, it encodes the analog of the evolution map $\psi_{\Li}\mapsto U_{[t_1,t_2]}\psi_{\Li}$, in the mixed state formalism, i.e., $\sigma_{\Li}\mapsto U_{[t_1,t_2]}\sigma_{\Li} U_{[t_1,t_2]}^\dagger$.

With definition \eqref{eq:ampltonp} the composition rule for null probes can be readily inferred from the composition rule for amplitudes, compare expression \eqref{eq:amplgluing}. Thus, if $M$ as a region is a union $M_1\cup M_2$, then \cite{Oe:dmf}
\be
\np_{M_1 \cup M_2}  (\sigma_{\Li} \otimes \sigma_{\Lf}) = \sum_{j \in J} \np_{M_1} (\sigma_{\Li} \otimes \xi_j) \np_{M_2} (\xi^*_j \otimes \sigma_{\Lf}) .
\label{eq:npgluing}
\ee
Here, $\{\xi_j\}_{j\in J}$ is an orthonormal basis of the space $\cB_{\Sigma}$\footnote{Note that the relevant inner product of $\cB_{\Sigma}$ is the Hilbert-Schmidt inner product.}, where $\Sigma$ is the hypersurface where $M_1$ and $M_2$ are glued together. Remarkably, this looks exactly like expression \eqref{eq:amplgluing}, except with pure state spaces replaced by mixed state spaces. Also analogous to the case of amplitudes, the composition of time-evolutions recovers the composition of quantum operations as maps that encode time evolution.

The composition rule for two arbitrary probes takes the very same form as that for the null probe. Suppose $A$ is a probe in $M_1$ and $B$ is a probe in $M_2$, then the composite probe $A\diamond B$ in $M_1\cup M_2$ is determined by,
\be
(A\diamond B)_{M_1 \cup M_2}  (\sigma_{\Li} \otimes \sigma_{\Lf}) = \sum_{j \in J} A_{M_1} (\sigma_{\Li} \otimes \xi_j) B_{M_2} (\xi^*_j \otimes \sigma_{\Lf}) .
\label{eq:pgluing}
\ee
What is more, if $A$ and $B$ are primitive probes then so is $A\diamond B$. This latter fact generalizes the fact in the SFQ that the composition of completely positive maps is completely positive.

It would be very suggestive to extend the correspondence between the amplitudes and the null probe to a correspondence between correlation functions and general probes. Indeed, the composition rule for the mixed state objects would follow in both cases from that of the pure state objects. However, while amplitudes and null probes have exactly the same physical meaning, this is not at all the case for correlation functions and probes. Nevertheless, we shall take advantage of the correspondence in composition laws as we advance.

\subsection{Discard and the Schwinger-Keldysh formalism}
\label{sec:SK}

The discard operation in SFQ consists in applying the trace to a state $\sigma$. The system is discarded at the time $t$ that the operation is applied. This implements a distinction between a "past", about which we might reason, and a "future" about which we may know nothing. In order to bring this into a spacetime setting we need a spacelike hypersurface representing the time $t$ and thus the division between "past" and "future". In the local PF we may formally associate the \emph{discard probe} to the region $[t,\infty)\times\R^3$ in the future of $t$, although all that really matters is the hypersurface at $t$. Thus, instead of $\discard_{[t,\infty)\times\R^3}$ we shall use the simplified notation $\discard_{t}$ for the map $\cB_t\to\R$, or even just $\discard$, if no ambiguity arises, given by the trace $\discard(\sigma)=\tr(\sigma)$.

We proceed to consider the composition of the discard probe with a null probe. To this end, we suppose that the latter corresponds to a time-interval region. At the end of the time-interval the system is discarded. This is,
\begin{equation}
    \discard_{t_2}\comp \np_{[t_1,t_2]}(\sigma)
    =\sum_{k\in I} \sum_{i\in I}
    \rho_{[t_1,t_2]} (\sigma\zeta_i\tens\zeta_k) \overline{\rho_{[t_1,t_2]}(\zeta_i\tens\zeta_k)} .
\end{equation}
We rewrite the amplitudes in terms of path integrals and also suppose, for simplicity, that the initial state is pure $\sigma=|\psi\rangle\langle\psi|$. Then,
\begin{align}
    \discard_{t_2}\comp \np_{[t_1,t_2]}(|\psi\rangle\langle\psi|)
    & =\sum_{k\in I}
    \int_{K_{[t_1,t_2]}}\xD\phi\, \psi(\phi|_1) \overline{\zeta_k(\phi|_2)}
    e^{\im S(\phi)} 
    \overline{\int_{K_{[t_1,t_2]}}\xD\phi'\, \psi(\phi'|_1) \overline{\zeta_k(\phi'|_2)} e^{\im S(\phi')}} \nonumber\\
    & =\int_{\substack{K_{[t_1,t_2]}^2\\ \phi|_2=\phi'|_2}}\xD\phi\,\xD\phi'\, \psi(\phi|_1)\overline{\psi(\phi'|_1)} 
    e^{\im S(\phi)-\im S(\phi')} .
    \label{eq:schkel1}
\end{align}
The last integral is a double path integral over field configurations $\phi$ and $\phi'$ in the time interval $[t_1,t_2]$ such that $\phi$ and $\phi'$ coincide at time $t_2$. This coincidence condition encodes the discard at time $t_2$. This double path integral is precisely that of the Schwinger-Keldysh in-in formalism \cite{Sch:brownianosc,Kel:diagramne}. We can think of $\phi$ as a field configuration that goes forward in time from $t_1$ to $t_2$ and then $\phi'$ as going backward in time from $t_2$ back to $t_1$. Due to unitarity of time evolution, we can shift the time $t_2$ arbitrarily, without changing the value of the integral, as long as it is not before the initial time $t_1$. In particular, if we shift $t_2$ to coincide with $t_1$, the path integral disappears,
\begin{equation}
    \discard_{t_2}\comp \np_{[t_1,t_2]}(|\psi\rangle\langle\psi|)
    =\discard_{t_1}(|\psi\rangle\langle\psi|) .
    \label{eq:npns}
\end{equation}

For the most part, considerations in the following will be limited to probes associated to spacetime regions that are time-intervals and settings where a discard is applied in the future. In that case it makes perfect sense to uphold the causality axiom, Figure~\ref{fig:causality_axiom}, left-hand side, and in this way specify what a non-selective probe in a time-interval region is. In particular, the null probe is non-selective in this way, as we have just seen expressed in equation \eqref{eq:npns}. Note that this also implies that denominators of expressions for probabilities or expectation values that consist only of a discard after the null probe applied to a normalized state take unit value and can therefore be ignored. We come back to the question of how the physical content of the causality axiom is generalized to the relativistic setting in Section~\ref{sec:relcausal}.


\section{Single measurements and observables}
\label{sec:single_meas}

\subsection{Single projective measurements in SFQ}

If we perform only a single measurement in the SFQ and subsequently discard the system, probabilities and expectation values take a particularly simple form. This is even more so, if the measurement is projective, i.e.\ the Kraus operators $\{K_i\}$ are projectors, compare expression \eqref{eq:kraus_decomp}. To emphasize that we deal with projectors, we shall write $\{P_i\}$. Thus, the probability for outcome $i$ in measuring an initial state $\sigma$ and discarding afterwards is simply $\tr(P_i\sigma P_i)=\tr(P_i\sigma)$. If we assign real values $a_i$ to the outcomes $i$, the expectation value is
\begin{equation}
    \langle A\rangle = \sum_i a_i \tr(P_i \sigma P_i)= \tr(A\sigma),\quad\text{with}\quad
    A=\sum_i a_i P_i .
\end{equation}
We then also say that we measure the expectation value of the self-adjoint operator $A$. We emphasize here that in this way we can calculate the expectation value of $A$ without explicitly knowing its spectral decomposition, i.e., the projectors $P_i$ and eigenvalues $a_i$. In case that the initial state is a pure state $\sigma=|\psi\rangle\langle\psi|$ we can even express the expectation value without any reference to a mixed state formalism,
\begin{equation}
    \langle A\rangle = \tr(A \sigma)=\tr(A|\psi\rangle\langle\psi|)=\langle\psi,A\psi\rangle .
    \label{eq:singleev}
\end{equation}

\subsection{Measuring single observables in QFT}
\label{sec:singleqft}

The fact that for a single measurement we only need the self-adjoint operator $A$ and not its spectral decomposition is particularly important in QFT. Since we have well-established procedures to construct the self-adjoint operators for important classes of classical observables, we can easily calculate the expectation values of these observables for some given initial state, discarding the system afterwards. What is more, instead of working explicitly with operators, we can work with the path integral quantization \eqref{eq:piquant} as reviewed in Section~\ref{sec:compqft}. Combining expressions \eqref{eq:piquant} and \eqref{eq:schkel1}, the expectation value of a classical observable $F$ localized in the time-interval $[t_1,t_2]$ and with initial pure state $\psi$ can be expressed through the Schwinger-Keldysh path integral,
\begin{equation}
   \langle F\rangle=\int_{\substack{K_{[t_1,t_2]}^2\\ \phi|_2=\phi'|_2}}\xD\phi\,\xD\phi'\, \psi(\phi|_1) \overline{\psi(\phi'|_1)} F(\phi) 
    e^{\im S(\phi)-\im S(\phi')} .
\end{equation}
Note that here we are no longer free (as in the case without observable) to move the initial and final boundaries of the path integral. Rather, the support of the observable $F$ has to be covered completely by the time-interval $[t_1,t_2]$. However, strictly speaking, this only applies to the forward path integral over $\phi$ and not the backward path integral over $\phi'$. Moving $t_1$ forward to $t_2$ for the backward path integral, this disappears, leaving an ordinary in-out path integral,
\begin{equation}
    \langle F\rangle=\int_{K_{[t_1,t_2]}}\xD\phi\, \psi(\phi|_1) \overline{\psi(\phi|_2)} F(\phi) 
     e^{\im S(\phi)} .
     \label{eq:singleqft}
\end{equation}
We have used the symbol $\psi$ for the initial state at $t_1$ as well as  its time-evolved version at $t_2$.


\section{The modulus-square construction}
\label{sec:modsquare}

In the SFQ we have a well-established compositional notion of measurement (as reviewed in Section~\ref{sec:sfq}) which can be extended to QFT through the local PF (Sections~\ref{sec:posform} and \ref{sec:probes_qft}) to a local and compositional notion of measurement. This clarifies the mathematical structures involved, generalizing \emph{quantum operations} to \emph{probes}. The remaining problem is the construction of those probes that correspond to specific classical observables. In the SFQ we usually have a quantization prescription that produces a self-adjoint operator from a real observable. In QFT it is convenient to use the path integral for the analog purpose, recall expression \eqref{eq:piquant}. Then, if we are only interested in the expectation value of a single measurement, there is the simple expression \eqref{eq:singleev} in the SFQ. The analog in QFT is the simple path integral \eqref{eq:singleqft}. However, if we want to compose measurements in SFQ, we need the full quantum operation corresponding to the observable. At the operator level it has recently been shown how this can be accomplished for linear observables in QFT via the spectral decomposition and in such a way that relativistic causality constraints are satisfied \cite{Oe:spectral}. However, we need a path integral "native" approach here to fulfill our aim of full spacetime compositionality. In the present section, we present such a proposal.

\subsection{Motivation}
\label{sec:msmot}

We start with the very simple observation that given any set of operators $\{K_i\}$, the map
\be
\sigma \mapsto  \Phi (\sigma) = \sum_i K_i \sigma K_i^\dag
\ee
is a completely positive map (compare Section~\ref{sec:sfq}). If it was the quantum operation implementing the measurement of an expectation value, then this expectation value for an initial state $\sigma$ would be,
\be
\tr(\sum_i K_i \sigma K_i^\dag)=\tr(\sum_i K_i^\dag K_i \sigma) .
\ee
Moreover, the corresponding self-adjoint operator would thus be,
\be
  A=\sum_i K_i^\dag K_i .
\ee

Interestingly, this very same scheme is replicated in the local PF using correlation functions of QFT. The key observation is that replacing the amplitude map in the definition \eqref{eq:ampltonp} of the null probe by a correlation function still yields a positive map $\cB_{\partial M}\to\R$, i.e., a map that may define a primitive probe. For simplicity, we consider a single real observable $F:K_M\to\R$ on a spacetime region $M$, corresponding to a single operator $K$ instead of a set $\{K_i\}$ as above. Recall that $K_M$ denotes here the space of field configurations in $M$. For later use it will be convenient to do our definitions with two possibly distinct observables $F$, $F'$. Define 
\be
 \cP_M[F | F'](\sigma)
 \defeq\sum_{k\in I} \rho_M[F](\sigma\zeta_k) \overline{\rho_M[F'](\zeta_k)} .
 \label{eq:msprobed}
\ee
Then, $\cP_M[F | F]$ is a positive map, i.e., a primitive probe.
If we let $M$ be a time-interval region $[t_1,t_2]$ and discard at time $t_2$ with an initial state $\sigma$ at time $t_1$, we obtain,
\begin{equation}
    \discard\comp \cP_{[t_1,t_2]}[F | F'](\sigma)
    =\sum_{k\in I} \sum_{j\in I} \rho_{[t_1,t_2]}[F] (\sigma\zeta_i\tens\zeta_k)
    \overline{\rho_{[t_1,t_2]}[F'](\zeta_i\tens\zeta_k)} .
\end{equation}
We may rewrite this as a Schwinger-Keldysh double path integral, as in Section~\ref{sec:SK}. To this end let $\sigma=|\psi\rangle\langle\psi|$ be a pure state,
\be
\discard\comp \cP_{[t_1,t_2]}[F | F'](|\psi\rangle\langle\psi|)
=\int_{\substack{K_{[t_1,t_2]}^2\\ \phi|_2=\phi'|_2}}\xD\phi\,\xD\phi'\, \psi(\phi|_1)\overline{\psi(\phi'|_1)} F(\phi) \overline{F'(\phi')}
    e^{\im S(\phi)-\im S(\phi')} .
\ee
To see the significance of this expression, we consider the operator analog with $F=F'$. Let $\hat{F}$ be the operator corresponding to the observable $F$ via Weyl quantization. Then we get,
\be
\discard\comp \cP_{[t_1,t_2]}[F | F](|\psi\rangle\langle\psi|)
=\tr(\hat{F} |\psi\rangle\langle\psi| \hat{F}^\dag)
=\tr(\hat{F}^\dag\hat{F} |\psi\rangle\langle\psi|) .
\ee
That is, we obtain the expectation value corresponding to the self-adjoint operator $\hat{F}^\dag\hat{F}$. We obtain a particular quantization of the classical observable
\be
  \Lambda=\overline{F} F = |F|^2 .
\ee
What is more, if $F$ is positive, then $\hat{F}^2=\widehat{F^2}=\hat{\Lambda}$. That is, we recover then exactly the expectation value of the quantization of the observable $\Lambda$. This encapsulates the essence of our present proposal. Since the observable $\Lambda$ is the modulus-square of the observable $F$, we call this the \emph{modulus-square construction}. This is intimately related to the modulus-square functor discussed in \cite{Oe:dmf,Oe:posfound}. As a generalization,
using linearity, we can replace $\cP[F | F]$ by $\sum_i \lambda_i \cP[F_i | F_i]$ (which is a not necessarily primitive probe) to encode a quantization of the classical observable $\Lambda=\sum_i \lambda_i |F_i|^2$. We will discuss an example of an even further generalization in Section~\ref{sec:emtens}.

\subsection{Proposal and properties}
\label{sec:properties}

Suppose we are given a classical observable $\Lambda:K_M\to\R$ in a spacetime region $M$, that takes only non-negative values. Let $F:K_M\to\R$ be the non-negative square-root. We then propose to take the primitive probe defined in equation \eqref{eq:msprobed} (with $F'=F$) as the probe that encodes the measurement of the expectation value of $\Lambda$. Moreover, we propose that the non-selective probe associated to this measurement is taken to be the null-probe. (Recall from Section~\ref{sec:posform} that probabilities or expectation values arise from quotients, where we also need to specify the non-selective probe corresponding to a given selective one.)
This proposal exhibits the following properties:

\paragraph{Positivity}
Positivity of probes is essential for a consistent probability interpretation in the PF. This is analogous to complete positivity of quantum operations in the SFQ, compare Section~\ref{sec:local_measurements}. This applies also to probes encoding expectation value measurements for positive observables. We have already remarked that the probe $\cP_M[F | F]$ is positive.

\paragraph{Locality}
Given an observable $\Lambda$ defined in a spacetime region $M$, the square-root $F$ is also defined in $M$. Then, locality of the path integral allows to define the probe $\cP_M[F | F]$ in the same spacetime region $M$. 

\paragraph{Compositionality}
The composition rule of correlation functions \eqref{eq:obsgluing} originating from the path integral induces a corresponding composition rule for probes. Thus, let $F_1$ and $F_2$ be classical observables in regions $M_1$ and $M_2$ respectively. Let $M=M_1\cup M_2$ be the joint region and $F=F_1 F_2$ the product observable. Then, combining \eqref{eq:obsgluing} and \eqref{eq:msprobed},
\begin{align}
\cP_M[F | F](\sigma_{\Li} \tens \sigma_{\Lf})
 & = (\cP_{M_1}[F_1|F_1]\diamond \cP_{M_2}[F_2 | F_2])_{M_1 \cup M_2}  (\sigma_{\Li} \otimes \sigma_{\Lf}) \nonumber\\
 & = \sum_{j \in J} \cP_{M_1}[F_1 | F_1] (\sigma_{\Li} \otimes \xi_j) \cP_{M_2}[F_2 | F_2] (\xi^*_j \otimes \sigma_{\Lf}) .
\end{align}
That is, the composition of the probes measuring $\Lambda_1= |F_1|^2$ and $\Lambda_2=| F_2|^2$ is the probe measuring $\Lambda=|F|^2=\Lambda_1 \Lambda_2= |F_1 F_2|^2$.

\paragraph{Single measurement recovery}
If we perform a single measurement and subsequently discard we want to recover the same expectation value as that obtained by the established standard methods of QFT, compare Section~\ref{sec:singleqft}. Indeed, we have demonstrated precisely this already in Section~\ref{sec:msmot}. Note that this comparison makes sense only in the case where $\Lambda$ encodes an instantaneous measurement, that is, if it depends on the field on a spacelike hypersurface only. (This is the notion of a \emph{slice observable} \cite{Oe:feynobs,CoOe:locgenvac}).

\subsection{Coherent states and factorization}

A further important element of our analysis of measurement in this work will be \emph{semiclassicality}. That is, to what extent can we reproduce classical expectation values for states that show a behavior close to classical solutions? These states are the \emph{coherent states}. In order to describe them adequately, we recall basic aspects of quantization of linear bosonic QFT. Thus, we start with the classical phase space $L$ that is equipped with a symplectic structure $\omega$. With the choice of vacuum comes a complex structure $J$ that combines with the symplectic structure into a complex inner product
\begin{equation}
\{\phi',\phi\}=g(\phi',\phi)+2\im\omega(\phi',\phi),
\end{equation}
where $g(\phi',\phi)=2\omega(\phi',J\phi)$.
This makes $L$ into a complex Hilbert space. The complex Hilbert space $\cH$ of pure states is then the \emph{Fock space} over $L$. Note that both $L$ and $\cH$ are specific to \emph{hypersurfaces}. In SFQ and standard treatments of QFT this is usually "forgotten" without consequence as some reference equal-time hypersurface is fixed. If we need to emphasize this, however, given a hypersurface $\Sigma$ we denote the associated phase space of germs of classical solutions by $L_{\Sigma}$ and the corresponding space of pure states by $\cH_{\Sigma}$.

Associated to each element of $\phi\in L$ is a \emph{normalized coherent state} $\ncoh_{\phi}\in\cH$. These have inner product,
\begin{equation}
  \langle \ncoh_{\xi},\ncoh_{\phi}\rangle=\exp\left(\frac12\{\phi,\xi\}-\frac14 \{\xi,\xi\}-\frac14 \{\phi,\phi\}\right) .
\end{equation}
The normalized coherent states also satisfy a \emph{completeness relation} with respect to a Gaussian measure $\nu$ on (an extension $\hat{L}$ of) $L$ as follows \cite{Oe:holomorphic},
\begin{equation}
  \langle \eta,\psi\rangle = \int_{\hat{L}}\xd\nu(\phi)\, \exp(\frac12\{\phi,\phi\})
   \langle \eta,\ncoh_{\phi}\rangle \langle\ncoh_{\phi},\psi\rangle .
   \label{eq:completeness}
\end{equation}
This completeness relation is also useful for mixed states. Introduce the operator $\mcoh_{\phi|\phi'}\defeq |\ncoh_{\phi}\rangle\langle\ncoh_{\phi'}|$. Note that this is only positive and thus a state if $\phi=\phi'$. In general, $\mcoh_{\phi|\phi'}^\dagger=\mcoh_{\phi'|\phi}$. Any state can be obtained as an integral over these particular generalized states, which is why we will use them in the following as if they were states,\footnote{The integral is to be understood weakly, i.e., with both sides evaluated with a norm-continuous function.}
\begin{equation}
  \sigma=\int_{\hat{L}\times\hat{L}}\xd\nu(\phi)\xd\nu(\phi')
  \exp(\frac12\{\phi,\phi\}+\frac12\{\phi',\phi'\}) \tr(\mcoh_{\phi'|\phi} \sigma)\, \mcoh_{\phi|\phi'} .
\end{equation}

Correlation functions take a particularly simple form on coherent states. To see this we start with a real linear observable $D:K_M\to\R$ and form the corresponding \emph{Weyl observable} $F\defeq \exp(\im D)$. We then have the \emph{factorization identity} \cite{Oe:feynobs,CoOe:locgenvac}, 
\begin{equation}
  \rho_M[F](\ncoh_\phi)=\rho_M(\ncoh_\phi) F(\phi^{\Lint}) \rho_M[F](\ncoh_0) .
  \label{eq:corfact}
\end{equation}
Note that $\ncoh_0$ is the vacuum state. Recall that $\phi$ is an element of $L_{\partial M}$, i.e., the classical phase space of germs of solutions on the boundary $\partial M$ of the region $M$. The complexification $L^\C_{\partial M}$ of $L_{\partial M}$ admits a direct sum decomposition $L_{\partial M}^\C=L_M^\C\oplus L_{X}^{\C}$, where $L_M$ encodes the solutions in the interior of $M$, and $L_X^\C$ encodes the vacuum boundary condition in the exterior $X$ of $M$ \cite{CoOe:vaclag}. This determines the decomposition $\phi=\phi^{\Lint}+\phi^{\Lext}$ with $\phi^{\Lint}\in L_M^\C$ and $\phi^{\Lext}\in L_X^{\C}$.
An important special case arises when the region $M$ is a time interval $[t_1,t_2]$ with the boundary the union of the hypersurface at $t_1$ and the hypersurface at $t_2$, $\partial M=\Sigma_1\cup \Sigma_2$. Then, a solution $\phi$ in a neighborhood of the boundary decomposes into a solution $\phi_1$ at $t_1$ and $\phi_2$ at $t_2$, $\phi=(\phi_1,\phi_2)$. The vacuum is then determined in the usual way by a decomposition into positive and negative energy solutions. In this case $\phi^{\Lint}=\phi_1^- + \phi_2^+$, where $\phi_1^-$ is the positive-energy component of $\phi_1$ and $\phi_2^+$ is the negative-energy component of $\phi_2$ \cite{CoOe:vaclag}.\footnote{Our sign conventions here are those of \cite{Oe:feynobs,CoOe:locgenvac} and opposite to those of \cite{CoOe:vaclag}.}

The vacuum correlation function $\rho_M[F](\ncoh_0)$ can be evaluated explicitly \cite{Oe:feynobs} in terms of the \emph{time-ordered observable propagator} $\wuu(D,D)$, see Appendix~\ref{sec:vacobs},
\begin{equation}
  \label{eq:vev}
  \rho_M[F](\ncoh_0)=\exp\left(-\frac12 \wuu(D,D)\right) .
\end{equation}
From the factorization identity of correlation functions \eqref{eq:corfact} follows a factorization identity for the probes \eqref{eq:msprobed}. Thus, let $D,D'$ be real linear observables $K_M\to\R$ and $F=\exp(\im D), F'=\exp(\im D')$ the corresponding Weyl observables. Then, with equations \eqref{eq:msprobed}, \eqref{eq:corfact} and \eqref{eq:vev},
\begin{equation}
   \cP_M[F | F'](\mcoh_{\phi|\phi'})=\np_M(\mcoh_{\phi|\phi'}) F(\phi^{\Lint}) \overline{F'(\phi'^{\Lint})} 
   \exp\left(-\frac12 \wuu(D,D) -\frac12 \overline{\wuu(D',D')}\right) .
   \label{eq:mspfact}
\end{equation}
Weyl observables in themselves are usually not of interest in measurement processes. Their usefulness lies in their role as generators of polynomial observables. Consider the product of linear observables $D_1\cdots D_n$. Define  $D_{\lambda}\defeq \lambda_1 D_1 + \cdots + \lambda_n D_n$, where $\lambda_1,\ldots,\lambda_n$ are real variables and $F_{\lambda}=\exp(\im D_{\lambda})$. Then,
\begin{equation}
  f(D_1\dots D_n)=(-\im)^n\frac{\partial}{\partial\lambda_1}\cdots \frac{\partial}{\partial\lambda_n} f(F_{\lambda}) \big|_{\lambda_1=\cdots=\lambda_n=0} ,
  \label{eq:functional_deriv_weyl}
\end{equation}
where $f$ is any linear functional on the space of observables. In particular, taking $f(F_{\lambda})$ to be $\rho[F_{\lambda}](\psi)$ yields $\rho[D_1\cdots D_n](\psi)$, i.e., the correlation function for a polynomial observable can be obtained in this way from the correlation function for a Weyl observable. This same procedure serves evidently in the case of modulus-square probes, using the factorization identity \eqref{eq:mspfact}.

\subsection{Semiclassicality and renormalization}
\label{sec:sclassrenorm}

Armed with the coherent states we consider the question of \emph{semiclassicality}. Starting from the premise that a coherent state $\ncoh_{\phi}$ approximates the associated classical solution $\phi$, we can demand that the quantum expectation value of measuring some observable should coincide with the classical expectation value. This is most straightforwardly achieved by normal-ordered quantization. However, normal-ordered quantization does not respect compositionality. That is, the composite of normal ordered observables is not in general the normal ordering of the composition of the observables. For compositionality to work we have to stick to Weyl quantization or its spacetime equivalent, path-integral quantization, see Section~\ref{sec:compqft}.
We therefore consider the question of semiclassicality for the simplest observables of interest here: quadratic observables, that is observables of the form $\Lambda= D^2$, where $D$ is real and linear.

There are two distinct scenarios: The first and simpler one is where we have a coherent state on the boundary of a region and make a measurement inside. Using \eqref{eq:mspfact} this yields for the modulus-square probe,
\begin{equation}
  \cP_M[D | D](\mcoh_{\phi})
  = \frac{\partial}{\partial\lambda} \frac{\partial}{\partial\lambda'}
  \cP_M[\exp(\im\lambda D)| \exp(\im\lambda' D)](\mcoh_{\phi}) \big|_{\lambda=\lambda'=0}
  =\np_M(\mcoh_{\phi}) |D(\phi^{\Lint})|^2 .
\end{equation}
We have used the simplified notation $\mcoh_{\phi}\defeq\mcoh_{\phi|\phi}$.
For the expectation value we obtain the quotient,
\begin{equation}
  \frac{\cP_M[D|D](\mcoh_{\phi})}{\np_M(\mcoh_{\phi})}
  =|D(\phi^{\Lint})|^2 .
\end{equation}
What is the interpretation of this? Recall that the solution $\phi$ is defined in a neighborhood of the boundary of $M$. $\phi^\Lint$ on the other hand is a complexified solution in the interior of $M$. In the classical theory, to talk about the expectation value of $\phi$ in $M$ can only make sense if $\phi$ is a solution in $M$. In other words, $\phi$ on the boundary needs to arise as the restriction of a solution in the interior. Precisely if this is the case, we have $\phi^{\Lint}=\phi$ in the quantum theory. Then, the quantum expectation value simply reduces to,
\begin{equation}
  |D(\phi^{\Lint})|^2=|D(\phi)|^2=\Lambda(\phi) .
\end{equation}
In other words, if the coherent state on the boundary is classically admissible, the quantum theory recovers the classical expectation value. And otherwise, it does not make sense to talk about a classical expectation value. The modulus-square probe achieves semiclassicality in this scenario. However, while this scenario is simple, it does not correspond to the standard measurement scenario, where the system is discarded after the measurement. We consider this in turn.

We take a time-interval region $M=[t_1,t_2]$. Note that the states we consider in the following are initial states rather than full boundary states as the role of the final state is taken by the discard operation.
The composition of the modulus-square probe with the discard yields (see Appendix~\ref{sec:derivations}),
\begin{equation}
  \label{eq:probe_discard}
  \discard\comp\cP_{[t_1,t_2]}[D|D](\mcoh_{\phi})
  =\Lambda(\phi)+ r(D,D) .
\end{equation}
Here, $r(D,D)=\Re(\wuu(D,D))$ is the real observable propagator, see Appendix~\ref{sec:vacobs}. On the other hand, the null probe yields,
\begin{equation}
  \discard\comp\np_{[t_1,t_2]}(\mcoh_{\phi})
  =\tr(\mcoh_{\phi})=\langle\ncoh_{\phi},\ncoh_{\phi}\rangle = 1 .
\end{equation}
For the expectation value we obtain the quotient, ($^\mathrm{ms}$ stands for "modulus-square")
\begin{equation}
  \langle\Lambda\rangle_{\mcoh_{\phi}}^{\mathrm{ms}}
  =\frac{\discard\comp\cP_{[t_1,t_2]}[D|D](\mcoh_{\phi})}{\discard\comp\np_{[t_1,t_2]}(\mcoh_{\phi})}
  =\Lambda(\phi)+ r(D,D) .
  \label{eq:sqeval}
\end{equation}
The semiclassically "correct" answer would be just $\Lambda(\phi)$. To achieve it, we introduce the \emph{renormalized} probe as follows,
\begin{equation}
  \cP_{[t_1,t_2]}^{\Lren}[D|D]\defeq \cP_{[t_1,t_2]}[D|D] - r(D,D) \np_{[t_1,t_2]} .
  \label{eq:renprobe}
\end{equation}
This yields, ($^\Lren$ stands for "renormalized modulus-square")
\begin{equation}
  \langle\Lambda\rangle_{\mcoh_{\phi}}^{\Lren}
  =\frac{\discard\comp\cP_{[t_1,t_2]}^{\Lren}[D|D](|\mcoh_{\phi})}{\discard\comp\np_{[t_1,t_2]}(\mcoh_{\phi})}
  =\Lambda(\phi) .
\end{equation}

To get a better understanding in which sense the subtraction just introduced is indeed a renormalization, we compare to the single measurement of a quadratic observable $\Lambda=D^2$ in the traditional quantum field theoretic scheme. For the comparison $D$ needs to be an instantaneous (slice) observable. Then, ($^{\mathrm{std}}$ stands for "standard")
\begin{equation}
  \langle\Lambda\rangle_{\mcoh_{\phi}}^{\mathrm{std}}
  =\langle\ncoh_\phi,\hat{\Lambda}\ncoh_\phi\rangle
  =\rho_{[t_1,t_2]}[\Lambda](\ncoh_\phi\tens\ncoh_\phi)
  =\Lambda(\phi)+\wuu(D,D) .
\end{equation}
Again, we are failing semiclassicality by an additive term. Indeed, we obtain a result identical to \eqref{eq:sqeval}. (Note that here $r(D,D)=\wuu(D,D)$ since $D$ is a slice observable.)
This is well known, and indeed the additive term $\wuu(D,D)$ is the Feynman propagator in the case of point-localized field observables and thus divergent, as will be discussed in Section~\ref{sec:pointsplit} and Appendix~\ref{sec:vacobs}. A standard way to subtract it is to use normal-ordered quantization. This restores semiclassicality and renormalizes the (potential) divergence,
\begin{equation}\label{semiclassical_lambda}
  \langle\no{\Lambda}\rangle_{\mcoh_{\phi}}^{\mathrm{std}}
  =\langle\ncoh_\phi,\no{\hat{\Lambda}}\ncoh_\phi\rangle
  =\rho_{[t_1,t_2]}[\no{\Lambda}](\ncoh_\phi\tens\ncoh_\phi)
  =\Lambda(\phi) .
\end{equation}
The subtraction \eqref{eq:renprobe} that we have introduced has an identical effect in the case of a single measurement, and removes the potentially divergent term $r(D,D)$. It does not in general amount to normal-ordering, however, in time-extended or composite measurements. Crucially, it removes the potential divergences without destroying compositionality, as we shall see.

\subsection{Renormalization and composition}
\label{sec:renormcompose}

As we have commented in Section~\ref{sec:sclassrenorm} expression~\eqref{eq:probe_discard}, the probe $\cP_{[t_1,t_2]}[D | D]$ does not yield the semiclassical expectation value on an initial coherent state when discarded. What is more, the second term $r(D,D)$ is divergent in cases of interest such as $D(\phi)=\phi(x)$ in scalar field theory, see Section~\ref{sec:pointsplit}. We have therefore introduced the \emph{renormalized probe} $\cP^\Lren_{[t_1,t_2]}[D | D]$ via expression~\eqref{eq:renprobe} with the term $r(D,D)$ removed. As a declared aim of our scheme is compositionality, it is of interest to understand how this renormalization behaves under composition. A priori, one would expect that the renormalization of a composite probe would not be simply related to the composition of renormalized versions of the component probes. Indeed, normal ordered quantization is precisely an example for this (see also the end of Section~\ref{sec:pointsplit}). In contrast, it turns out that in our scheme, the renormalization of a composite of quadratic probes is precisely the composition of the renormalized quadratic probes.

Thus, consider linear observables $D_1$ and $D_2$, with spacetime support in time intervals $[t_1,t]$ and $[t,t_2]$ respectively.\footnote{We could easily generalize the condition from requiring consecutive support to merely requiring non-overlapping support in spacetime. For simplicity of presentation we refrain from doing so here. Nevertheless, the result we obtain is fully general.} Evaluating the corresponding modulus-square probe on the generalized state $\mcoh_{\phi|\phi'}$ and discarding yields from expression \eqref{eq:dprobe} of Appendix~\ref{sec:quadobs},
\begin{multline}
    \discard\comp\cP_{[t_1,t_2]}[D_1 D_2 | D_1 D_2](\mcoh_{\phi|\phi'})
    =\discard\comp\np_{[t_1,t_2]}(\mcoh_{\phi|\phi'})
    \Big(
      2 |\wuu (D_1,D_2)|^2\\
      + r(D_1 ,D_1)  r(D_2 ,D_2)
      + 4r(D_1,D_2)
      \wD_1
      \wD_2
      +r(D_1,D_1) 
      \wD_2 
      \wD_2
      +r(D_2,D_2)
      \wD_1 
      \wD_1
      +\wD_1^2
      \wD_2^2
    \Big) .
    \label{eq:dprobeeval}
\end{multline}
Here, we have used the abbreviated notation $\widetilde{D}=D(\phi^{-}+\phi'^{+})$. Also, we have used the identity $\wuu(D_1,D_2)=\wud(D_1,D_2)$ due to the temporal ordering of $D_1$ before $D_2$. The potentially divergent terms arise from the evaluation of observable propagators on the same observable. These are the terms containing $r(D_1,D_1)$, $r(D_2,D_2)$ or both. Concretely, in the parenthesis this is the second, fourth and fifth term, which we need to remove to renormalize.

Instead, we consider now the composition of renormalized probes \eqref{eq:renprobe} for the same observables $\Lambda_1=D_1^2$ and $\Lambda_2=D_2^2$. This yields,
\begin{multline}
    \discard\comp\cP_{[t,t_2]}^{\Lren}[D_2 | D_2]\comp\cP_{[t_1,t]}^{\Lren}[D_1 | D_1](\mcoh_{\phi|\phi'}) \\
    =\discard\comp\left(\left(\cP_{[t,t_2]}[D_2 | D_2]-r(D_2,D_2)\np_{[t,t_2]}\right)\comp\left(\cP_{[t_1,t]}[D_1 | D_1]-r(D_1,D_1)\np_{[t_1,t]}\right)\right)(\mcoh_{\phi|\phi'}) \\
    =\discard\comp\left(\cP_{[t_1,t_2]}[D_1 D_2 | D_1 D_2]
      -r(D_2,D_2)\cP_{[t_1,t_2]}[D_1 | D_1]
      -r(D_1,D_1)\cP_{[t_1,t_2]}[D_2 | D_2] \right. \\
      \left. +r(D_2,D_2) r(D_1,D_1)\np_{[t_1,t_2]}
      \right)(\mcoh_{\phi|\phi'}) \\
    =\discard\comp\np_{[t_1,t_2]}(\mcoh_{\phi|\phi'})
    \Big(
      2 |\wuu (D_1,D_2)|^2
      + 4 r(D_1,D_2)
      \wD_1
      \wD_2
      +\wD_1^2
      \wD_2^2
    \Big) .
    \label{eq:reno_comp}
\end{multline}
Remarkably, we obtain precisely the renormalized version of expression~\eqref{eq:dprobeeval}, with the potentially divergent terms removed. Indeed, it is easy to see that this works not just for the composition of two, but of any number of renormalized modulus-square probes of quadratic observables. That is, "renormalization commutes with composition".


\section{Scalar field theory}
\label{sec:scalar}

After presenting the formalism for probe maps and the modulus-square approach in general bosonic linear field theory in globally hyperbolic spacetime, we specialize to Klein-Gordon theory. We focus on probes derived from field observables or linear functions of these and on the case in which the system is discarded at the end of the measurement(s). We also limit ourselves to the case of time-interval regions for simplicity and ease of comparison with other approaches. Time-interval regions are understood with respect to some global time function, although the existence of such a function is not essential, and we might as well consider regions delimited by arbitrary pairs of non-intersecting Cauchy hypersurfaces. What is essential though is that we assume the choice of a global vacuum state. (In particular, in- and out-vacuum are assumed the same.) We omit the explicit mention of the specific time-intervals when this does not introduce ambiguity.

\subsection{Local field operators and point-splitting}
\label{sec:pointsplit}

Our first example is measuring the square of the scalar field at a point. This serves to exhibit the divergences alluded to in Sections~\ref{sec:sclassrenorm} and \ref{sec:renormcompose} and justifies their removal as in Definition~\eqref{eq:renprobe} through point-splitting regularization. Thus, we consider the linear classical observable $D_x (\phi):= \phi (x)$. Instead of directly evaluating the corresponding modulus-square probe as in expression \eqref{eq:probe_discard}, we allow different evaluation points $x, x'$ in the forward- and backward-branch of the Schwinger-Keldysh path integral using expression \eqref{eq:probe_quadr_bare},\footnote{Another possible regularization scheme would consist of a family $D_x^\epsilon$ of suitably smeared non-singular observables converging to $\phi(x)$ for $\epsilon\to 0$. Then the corresponding probe could be taken with this same observable in both branches, $\cP[D_x^\epsilon|D_x^\epsilon]$, taking the limit after subtraction.}
\bal\label{field_quadratic}
\discard \diamond\cP[D_{x} | D_{x'}] (\mcoh_{\phi}) = 
\phi(x) \phi(x') +\wtud (x, x') .
\eal
The first term is the classical value of the observable (when $x=x'$) associated with the phase space element $\phi$ of the coherent state. The second term is the Wightman propagator (see also Appendix~\ref{sec:vacobs}) which is independent of the quantum state and contains information about the short-distance behavior of the field.
At the coincidence limit $x' \rightarrow x$ this term is the divergent vacuum expectation value,
\bal
\discard \diamond\cP[D_{x} | D_{x'}] (\mcoh_0) =\wtud (x, x') .
\eal
At $x'\neq x$ we can subtract this divergent term, justifying in the present case the definition of the renormalized probe \eqref{eq:renprobe}.
\bal\label{field_quadratic_ren}
\discard \diamond\cP^\Lren[D_{x} | D_{x}] (\mcoh_{\phi}) \defeq \lim_{x'\to x}
\discard \diamond\cP^\Lren[D_{x} | D_{x'}] (\mcoh_{\phi})
=\phi^2(x) .
\eal
Eq.~\eqref{field_quadratic} coincides with the result we obtain from the standard in-out formalism which is usually applied to single measurements \cite{Ful:qftcurved,Wal:qftcurved} as expected, recall Section~\ref{sec:single_meas}.

However, we now go beyond the capabilities of the standard formalism in describing joint measurements at different spacetime points in a way that not only respects locality, but is also fully compositional.
As a simple example we consider the measurement of the composite observable $\phi^2(x_1) \phi^2(x_2)$ with $x_1\neq x_2$. We model this as a composition of two renormalized modulus-square probes following the procedure as described in Section~\ref{sec:renormcompose}. However, we modify this procedure by again implementing an explicit point-splitting regularization. Thus, using \eqref{eq:dprobe} we find for the unrenormalized point-split expectation value,
\begin{multline}
\label{eq:fieldcomppprobe}
\discard\comp \cP [D_{x_{1}} D_{x_{2}} | D_{x'_{1}} D_{x'_{2}}] (\mcoh_{\phi}) =
\wtud (x_1,x_2')\wtud (x_2,x_1') + \wtud (x_{1} , x_1') \wtud (x_{2} , x_{2}') \\
+\wtuu (x_1,x_2) \wtdd (x_1',x_2')
+ \wtud (x_1 , x_1')\phi (x_{2}) \phi (x_{2}') 
+ \wtud (x_{2}, x_{2}')\phi (x_{1}) \phi (x_{1}')
+\wtud(x_1,x_2')\phi(x_2) \phi(x_1') \\
+\wtud (x_2,x_1') \phi (x_1) \phi (x_2') 
+ \wtuu (x_1,x_2) \phi(x_1') \phi (x_2')
+ \wtdd (x_1',x_2') \phi (x_1) \phi (x_2)
+\phi (x_{1}) \phi (x_1')\phi (x_{2}) \phi (x_{2}') .
\end{multline}
The terms divergent in the limit $x_i' \rightarrow x_i$ are the terms containing the Wightman propagators $\wtud (x_1 , x_1')$ and $\wtud (x_{2}, x_{2}')$. Removing these and taking the coincidence limit $x_i'\to x_i$ yields the same result as composing the renormalized probes,
\begin{multline}
 \discard\comp \cP^\Lren [D_{x_2} | D_{x_2}]\comp\cP^\Lren[D_{x_1}| D_{x_1}] (\mcoh_{\phi}) \\
 = \phi^2 (x_1)\phi^2 (x_2) 
+2 \abs{\wtuu (x_1,x_2)}^2  
+ 4 \rp(x_1,x_2) \phi (x_1)\phi (x_2).
\end{multline}
(Note that we have used identities such as $|\wtuu(x_1,x_2)|=|\wtud(x_1,x_2)|$ etc.) Thus, we see again that renormalization "commutes" with composition, compare Section~\ref{sec:renormcompose}. The special case of the vacuum expectation value is,
\bal
 \discard\comp \cP^\Lren [D_{x_2} | D_{x_2}]\comp\cP^\Lren[D_{x_1}| D_{x_1}] (\mcoh_{0})
 = 2 \abs{\wtuu (x_1,x_2)}^2 .
\eal
This shows clearly the difference to a (non-compositional) normal-ordering prescription for the total probe, where the result would have been zero.

\subsection{The energy-momentum tensor}
\label{sec:emtens}

An important observable in physical theories is the energy-momentum tensor which in scalar field theory is given by
\bal\label{eq:emtensor}
T_{\mu\nu} (x)[\phi]=(\partial_{\mu}\phi(x)) (\partial_{\nu}\phi(x))
-\frac12 g_{\mu \nu}(x) (\partial_{\rho}\phi(x)) (\partial^{\rho}\phi(x)) + \frac12 m^2 g_{\mu \nu}(x) \phi^2(x)
\eal
We use the square bracket to explicitly indicate on which field, here $\phi$, the energy-momentum tensor is evaluated. In this way we may think of $T_{\mu\nu} (x)$ as a functional on field configurations (or solutions). In order to take advantage of point-splitting regularization we rewrite it as a coincidence limit,
\bal
T_{\mu\nu} (x)[\phi]= \lim_{x' \rightarrow x}
\Theta_{\mu\nu} (x,x')\phi (x)\phi (x'),
\eal
where we have defined the bi-differential operator
\bal
\Theta_{\mu\nu}(x,x')\defeq
&\frac{1}{2}\left(\partial_{\mu} \partial_{\nu}'  + \partial_{\nu}  \partial_{\mu}' \right)
-\frac{1}{4} (g_{\mu \nu}(x)+g_{\mu \nu}(x')) \partial_{\rho} \partial'^{\rho} + \frac{1}{4} m^2 (g_{\mu \nu}(x)+g_{\mu \nu}(x')) .
\eal
With $D_x(\phi)\defeq \phi(x)$ we could define the energy-momentum probe by the expression
\bal \label{eq:energyprobe}
 \cP [T_{\mu\nu}(x)]  \defeq
\lim_{x' \rightarrow x} \Theta_{\mu\nu}(x,x')\cP [ D_x | D_{x'}] .
\eal
However, we already know that this form of the field probe is in general divergent, but that we can fix it by renormalization (see Section~\ref{sec:pointsplit}). This motivates the definition of the renormalized energy-momentum probe map,
\bal \label{eq:energyproberen}
 \cP^\Lren [T_{\mu\nu}(x)] \defeq
\lim_{x' \rightarrow x} \Theta_{\mu\nu}(x,x')\cP^\Lren [ D_x | D_{x'}] .
\eal
Measuring the energy-momentum tensor on a pure coherent state then yields the semiclassical expectation value,
\bal
\discard\comp\cP^\Lren [T_{\mu\nu}(x)] (\mcoh_{\phi})
&=\lim_{x' \rightarrow x} \Theta_{\mu\nu}(x,x') \discard\comp\cP^\Lren[ D_x | D_{x'}](\mcoh_{\phi}) \nonumber\\
&=\lim_{x' \rightarrow x} \Theta_{\mu\nu}(x,x')
\Big( 
\phi (x) \phi (x')
\Big)
=T_{\mu\nu}(x)[\phi] .
\eal

As an example of a compositional observable we can consider the correlation function of the energy-momentum tensor at two different spacetime points. That is, we want to measure the expectation value of the product $\phi\mapsto T_{\mu_1\nu_1}(x_1) T_{\mu_2\nu_2}(x_2) [\phi]$. Our emphasis here is not on the physical interest in the particular quantity calculated, but on a proof of concept demonstration for measurement through the composition of probes, satisfying all the postulated properties.
Using the renormalized probes, this is,
\bal
&\discard\comp\cP^\Lren [T_{\mu_2\nu_2}(x_2)]\comp\cP^\Lren [T_{\mu_1\nu_1}(x_1)] (\mcoh_{\phi}) \nonumber\\
& =
\lim_{x_2' \rightarrow x_2}\lim_{x_1' \rightarrow x_1}
\Theta_{\mu_2 \nu_2}(x_2,x_2')\Theta_{\mu_1 \nu_1}(x_1,x_1')
\discard\comp\cP^\Lren[ D_{x_{2}} |  D_{x'_{2}}]
\comp\cP^\Lren[D_{x_{1}} | D_{x'_{1}} ](\mcoh_\phi) \nonumber\\
& =\lim_{x_2' \rightarrow x_2}\lim_{x_1' \rightarrow x_1}
\Theta_{\mu_2 \nu_2}(x_2,x_2')\Theta_{\mu_1 \nu_1}(x_1,x_1')
\Big(\phi (x_{1}) \phi (x_1')\phi (x_{2}) \phi (x_{2}') \nonumber\\
&\quad +\wtud (x_1,x_2')\wtud (x_2,x_1')
+\wtuu (x_1,x_2) \wtdd (x_1',x_2')
+\wtud(x_1,x_2')\phi(x_2) \phi(x_1') \nonumber\\
&\quad +\wtud (x_2,x_1') \phi (x_1) \phi (x_2') 
+ \wtuu (x_1,x_2) \phi(x_1') \phi (x_2')
+ \wtdd (x_1',x_2') \phi (x_1) \phi (x_2) \Big) \nonumber\\
& =\lim_{x_2' \rightarrow x_2}\lim_{x_1' \rightarrow x_1}
\Theta_{\mu_2 \nu_2}(x_2,x_2')\Theta_{\mu_1 \nu_1}(x_1,x_1')
\Big(\phi (x_{1}) \phi (x_1')\phi (x_{2}) \phi (x_{2}') \nonumber\\
&\quad +2\wtuu (x_1',x_2')\wtdd (x_1,x_2)
+4\rp(x_1',x_2')\phi(x_1) \phi(x_2) \Big) .
\eal
Note that starting from expression~\eqref{eq:fieldcomppprobe} we have used identities such as $\wtud(x_1,x_2)\wtud(x_2,x_1)=\wtuu(x_1,x_2)\wtdd(x_1,x_2)$ (see Appendix~\ref{sec:vacobs}) as well as symmetries under exchange of variables $x_i \rightleftarrows x_i'$ and indices. Evaluating the bi-differential operators we obtain the following expression,
\bal\label{eq:en-mom_comp}
&\discard\comp\cP^\Lren [T_{\mu_2\nu_2}(x_2)]\comp\cP^\Lren [T_{\mu_1\nu_1}(x_1)] (\mcoh_{\phi})
= 
T_{\mu_1\nu_1}(x_1) T_{\mu_2\nu_2}(x_2) [\phi] 
\nonumber\\& 
+\Re\left(\wtuu_{\mu_1 \mu_2}(x_1,x_2)\wtdd_{\nu_1 \nu_2}(x_1,x_2)\right) +\Re\left(\wtuu_{\mu_1 \nu_2}(x_1,x_2)\wtdd_{\nu_1 \mu_2}(x_1,x_2)\right) 
\nonumber\\& 
+\rp_{\nu_1 \nu_2}(x_1,x_2)\phi_{\mu_1} (x_1) \phi_{\mu_2} (x_2) +\rp_{\mu_1 \mu_2}(x_1,x_2)\phi_{\nu_1} (x_1) \phi_{\nu_2} (x_2) 
\nonumber\\& 
+\rp_{\mu_1 \nu_2}(x_1,x_2)\phi_{\nu_1} (x_1) \phi_{\mu_2} (x_2)
+\rp_{\nu_1 \mu_2}(x_1,x_2)\phi_{\mu_1} (x_1) \phi_{\nu_2} (x_2)
\nonumber\\&
- g_{\mu_1 \nu_1}(x_1) \Big( \rp_{\rho_1\mu_2}(x_1,x_2) \phi^{\rho_1} (x_1) \phi_{\nu_2} (x_2) + \rp_{\rho_1\nu_2}(x_1,x_2) \phi^{\rho_1} (x_1) \phi_{\mu_2} (x_2)\Big)
\nonumber\\&
- g_{\mu_2 \nu_2}(x_2) \Big(\rp_{\mu_1\rho_2}(x_1,x_2) \phi^{\rho_2} (x_2) \phi_{\nu_1} (x_1) + \rp_{\nu_1\rho_2}(x_1,x_2) \phi^{\rho_2} (x_2) \phi_{\mu_1} (x_1)\Big)
\nonumber\\&
+m^2 g_{\mu_1 \nu_1}(x_1) \Big(\rp_{\mu_2}(x_2,x_1) \phi_{\nu_2} (x_2)+ \rp_{\nu_2}(x_2,x_1) \phi_{\mu_2} (x_2)\Big)\phi(x_1) \nonumber\\
& +m^2 g_{\mu_2 \nu_2}(x_2) \Big(\rp_{\mu_1}(x_1,x_2) \phi_{\nu_1} (x_1)+ \rp_{\nu_1}(x_1,x_2) \phi_{\mu_1} (x_1)\Big)\phi(x_2)
\nonumber\\&
+\frac{1}{4} g_{\mu_1 \nu_1}(x_1)g_{\mu_2 \nu_2}(x_2) \Big(2 \wtuu_{\rho_1 \rho_2} (x_1,x_2) \wtdd^{\rho_1 \rho_2} (x_1,x_2) + 4 \rp_{\rho_1 \rho_2} (x_1,x_2)\phi^{\rho_1} (x_1) \phi^{\rho_2} (x_2)\Big)
\nonumber\\&
-\frac{m^2}{2} g_{\mu_1 \nu_1}(x_1)g_{\mu_2 \nu_2}(x_2) 
\left(\Re(\wtuu_{\rho_2} (x_2,x_1) \wtdd^{\rho_2} (x_2,x_1)) + 2 \rp_{\rho_2}(x_2,x_1) \phi^{\rho_2} (x_2) \phi (x_1) \right.
\nonumber\\&
\left. +\Re(\wtuu_{\rho_1} (x_1,x_2) \wtdd^{\rho_1} (x_1,x_2)) + 2 \rp_{\rho_1}(x_1,x_2) \phi^{\rho_1} (x_1) \phi (x_2) 
\right)
\nonumber\\&
+\frac{m^4}{4} g_{\mu_1\nu_1} (x_1)g_{\mu_2\nu_2} (x_2)
\left( 2 \abs{\wtuu (x_1,x_2)}^2  
+ 
4 \rp (x_1,x_2) \phi (x_1)\phi (x_2)\right)
\nonumber\\&
-g_{\mu_1 \nu_1}(x_1) \left(
\Re(\wtuu_{\rho_1\mu_2} (x_1,x_2) {\wtdd^{\rho_1}}_{\nu_2} (x_1,x_2))
-m^2 \Re(\wtuu_{\mu_2} (x_2,x_1) \wtdd_{\nu_2} (x_2,x_1))
\right)
\nonumber\\&
-g_{\mu_2 \nu_2}(x_2) \left(
\Re(\wtuu_{\mu_1 \rho_2} (x_1,x_2) {\wtdd_{\nu_1}}^{\rho_2} (x_1,x_2))
-m^2 \Re(\wtuu_{\mu_1} (x_1,x_2) \wtdd_{\nu_1} (x_1,x_2))
\right) .
\eal
We have made use of the following notation for the derivatives 
\bal
h_{\rho}(x)=\frac{\partial}{\partial x^{\rho}} h(x),\quad
h_{\rho} (x_1,x_2)\defeq \frac{\partial}{\partial x_1^{\rho}} h(x_1,x_2),\quad
h_{\rho \rho'} (x_1,x_2)\defeq \frac{\partial^2}{\partial x_1^{\rho}\partial x_2^{\rho'}} h(x_1,x_2) .
\eal
The result is the real and finite correlation function for the measurement of the product of specified components of the energy-momentum tensor at two different spacetime locations in our approach. Note in particular, that by construction the result is completely symmetric under the exchange $(x_1,\mu_1,\nu_1)\leftrightarrow (x_2,\mu_2,\nu_2)$, in spite of the initial use (for convenience) of consecutive time intervals in its construction.


\subsection{Relativistic causality}
\label{sec:relcausal}

\begin{figure}[ht]
    \centering
\includegraphics[scale=0.7]{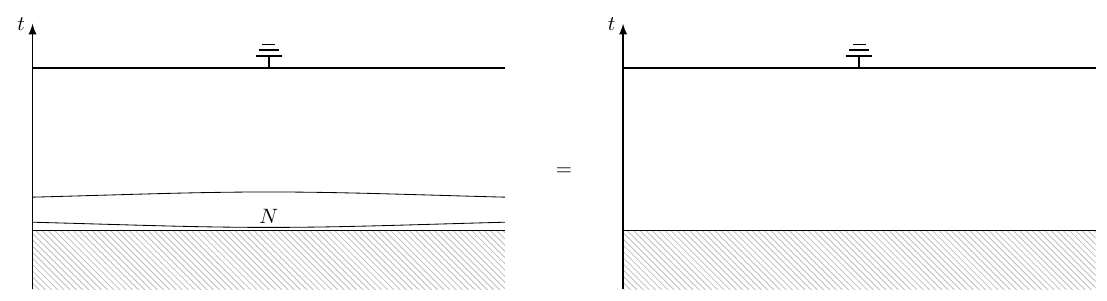}
\caption{Spacetime representation of the non-relativistic causality axiom. The axiom involves a non-selective probe at $N$. The shaded region may contain other probes. It comprises the part of the spacetime that is not in the future of $N$ (i.e.\ it's past).}
\label{fig:stcausal}
\end{figure}

\begin{figure}
    \centering
\includegraphics[scale=0.7]{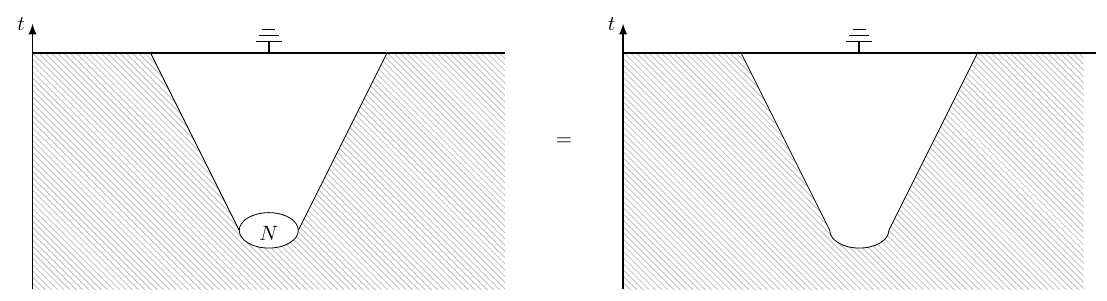}
\caption{Spacetime representation of the relativistic causality axiom. The axiom involves a non-selective probe at $N$. The shaded region may contain other probes. It comprises the part of spacetime that is not in the \emph{causal} future of $N$.}
\label{fig:relcausal}
\end{figure}

\begin{figure}
    \centering
\includegraphics[scale=0.7]{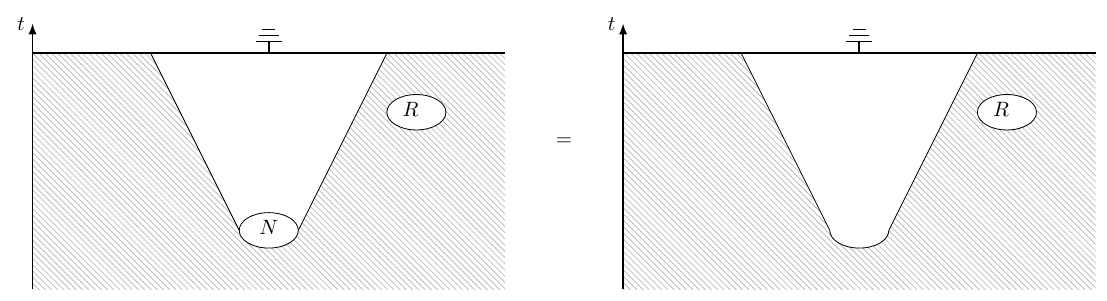}
\caption{Setup as in Figure~\ref{fig:relcausal}, but in addition to the non-selective probe at $N$, a selective probe at $R$ is singled out.}
\label{fig:locality}
\end{figure}

The \emph{causality axiom} of the SFQ finds a natural generalization in the relativistic context of QFT. We recall that the former is implemented by requiring that any non-selective probe composed with the discard in the future is equal to the discard alone, see Figure~\ref{fig:causality_axiom} (left-hand side). Crucially, we assume there that no other measurement is taking place in the future of our probe measurement, that is, until the discard. On the other hand, any other measurement may take place before. For a spacetime representation of the axiom, see Figure~\ref{fig:stcausal}. The generalization to the relativistic setting is now straightforward. Instead of disallowing other measurements in all of the future of our non-selective measurement, we disallow them only in the \emph{causal} future, since signals may not travel faster than light. This is illustrated in Figure~\ref{fig:relcausal}. We may also generate the relativistic condition by considering all possible Lorentz boosts of the non-relativistic condition.

Due to the restricted way composition works in SFQ, namely exclusively via past inputs and future outputs, it is possible to impose the causality axiom as a property of quantum operations, as we have recalled in Section~\ref{sec:sfq}. The situation is more involved for probes in the relativistic setting. It is clear that a non-selective probe that is embedded in a time-interval region has to satisfy the same composition identity with the discard as in the non-relativistic setting, compare our previous comments at the end of Section~\ref{sec:SK}. In general, however, the relativistic causality axiom can only be expressed as a condition on certain compositions of probes and their underlying spacetime regions.

\emph{Relativistic causality} goes hand in hand with \emph{spacetime locality}, i.e., the idea that a measurement is restricted to a certain spacetime region which may be specified explicitly. We have made this evident in Figure~\ref{fig:relcausal} for the non-selective probe by assigning it an extended spacetime region. To disentangle the two notions we insist, as in the non-relativistic case of SFQ, that the (relativistic) causality identity is an exclusive property of non-selective probes, namely those satisfying the identity depicted in Figure~\ref{fig:relcausal}. Of course, as in SFQ, we may then say that any (possibly selective) measurement for which this probe encodes the non-selective version, satisfies the (relativistic) causality condition. On the other hand we say that a probe is \emph{local} to a \emph{spacetime region} $R$ if for any composition of the probe with other probes and a non-selective probe $N$ satisfying the relativistic causality condition, such that $R$ and the regions of the other probes do not intersect the causal future of $N$, the identity of Figure~\ref{fig:locality} holds. Compared to Figure~\ref{fig:relcausal} we have here added explicitly a selective probe in region $R$ and the identity is then understood as encoding the locality property of this probe.

The modulus-square probes in the present setting encode expectation values and not directly probabilities. In particular, we have not constructed a corresponding non-selective probe. Indeed, we have proposed to use the null probe to take the role of the non-selective probe, which seems to work well in the contexts we have considered. The null probe of course satisfies trivially the (relativistic) causality property. Thus, for general modulus-square probes it only makes sense to check spacetime locality in the sense of our definition above. On the other hand, we can construct a particular non-trivial modulus-square probe that is non-selective in the usual sense. Given a linear observable $D_1$ we define $F=\exp(\im D_1)$ and consider the probe $\cP[F|F]$. For $D_1(\phi)=-\lambda\phi(x_1)$ this is exactly the "kick" as already used by Sorkin~\cite{Sor:impossible}. We use this to exemplify the relativistic causality axiom, with the other probe taken to be $\cP[D_2|D_2]$ for the linear observable $D_2(\phi)=\phi(x_2)$, or rather its renormalized version $\cP^{\Lren}[D_2|D_2]$. For simplicity, we take the initial state to be a coherent state.

The relevant expression is,
\bal
& \discard \comp \cP^{\Lren}[D_2|D_2]\comp \cP[e^{-\im \lambda D_{1}} |e^{-\im \lambda D_{1}}] (\mcoh_{\phi}) \nonumber\\
& =\discard \comp \cP[e^{-\im \lambda D_{1}} D_2 |e^{-\im \lambda D_{1}} D_2] (\mcoh_{\phi})
-\wtud(x_1,x_1)\discard \comp \cP[e^{-\im \lambda D_{1}} |e^{-\im \lambda D_{1}}] (\mcoh_{\phi}) \\
& = \discard(\mcoh_\phi) 
\left(\left(\im \phi(x_2) -\lambda (\wtud(x_2,x_1)- \wtuu(x_1,x_2))\right)
\left(-\im\phi(x_2) - \lambda (\wtud(x_1,x_2) - \wtdd(x_1,x_2))\right)
\right) . \nonumber
\eal
In the second line, we have sloppily left out the point-splitting regularization, but the reader can fill this in immediately now. We have also used the identity \eqref{eq:pid3} from Appendix~\ref{sec:quadobs}.

As for the $\lambda$-dependent terms we note that the combination
\be
\wtud(x_2,x_1) - \wtuu(x_1,x_2)=\overline{\wtud(x_1,x_2) - \wtdd(x_1,x_2)}
\ee
vanishes precisely if $x_2$ is not in the causal future of $x_1$. That is, the result in this case is precisely the same as if the non-selective probe $\cP[F|F]$ was not there. As expected, the identity depicted in Figure~\ref{fig:locality} is satisfied. What is more, we can see that the measurement at $x_2$ generically does detect the "kick" at $x_1$ if $x_2$ is in the causal future of $x_1$.


\section{Discussion and Outlook}

\label{sec:discussion}

In this work we have proposed a method for describing the joint measurement (in the sense of joint expectation) of an important class of observables in quantum field theory (QFT). The underlying \emph{modulus-square} construction (Section~\ref{sec:modsquare}) implements a \emph{quantization prescription} in the sense of providing objects of the quantum theory corresponding to classical observables. In contrast to what is usually understood by a quantization prescription, these objects are neither operators on Hilbert space nor elements of other types of algebras. Rather, they are \emph{spacetime probes}, the latter being a direct generalization of the standard notion of \emph{quantum operation}.\footnote{Note that in some parts of the literature the word \emph{probe} has been used in a different sense, namely as designating a system representing an ancilla or apparatus.} Crucially, this allows for a description of composite measurements through a corresponding composition of probes. This is analogous to the temporal composition of quantum operations, but generalizes this composition to a fully spacetime-local notion.
At the conceptual level, the underlying framework is the local positive formalism (PF), see Section~\ref{sec:posform}, combining spacetime locality and compositionality with a coherent probabilistic interpretation, generalizing that of the standard formulation of quantum theory (Section~\ref{sec:sfq}). At the technical level we take advantage of the path integral and its spacetime compositionality properties for amplitudes and correlation functions of observables, connecting with the Schwinger-Keldysh formalism in the common situation of a discard of the system in the future (Section~\ref{sec:probes_qft}). As a consequence, in the \emph{modulus-square} construction the probes inherit the spacetime locality properties of the original classical observables, seen as functions on spacetime field configurations. In particular, we show that a measurement constructed in this way cannot detect a non-selective "kick" measurement that is outside the causal past of the spacetime region supporting the observable (Section~\ref{sec:relcausal}). In this sense it satisfies relativistic causality.

We emphasize that in the presented construction it makes a priori no sense to ask for the measurement of a given self-adjoint operator. Rather, from the presented point of view the self-adjoint operators of the standard measurement scheme are merely intermediate objects, obtained by "conventional" quantization from a classical observable on phase space. They are intermediate because they are themselves then the source for the construction of quantum operations via the spectral decomposition. This intermediate operator stage does not exist in this form in the present construction. Rather, the starting point are classical observables on field configuration space and the construction proceeds via the path integral and the Schwinger-Keldysh formalism. If one wants to insist on taking a self-adjoint operator as the starting point for our construction, one would first have to find a classical phase space observable which yields this operator via quantization, promote this classical observable from phase space to field configuration space and then apply our construction. While this might be possible in many cases, contemplating this question is beyond the scope of the present paper.

The modulus-square construction allows quantizing observables that can be represented as squares of simpler observables such as field operators and their composites. This originates from the observation that a positivity constraint on probes can be satisfied by combining the correlation function for an observable with its complex conjugate (Section~\ref{sec:complocpf}), generalizing the construction of quantum operations from Kraus operators in the standard formulation of quantum theory (SFQ). The positivity constraint is analogous to complete positivity in the SFQ and plays a crucial role in the consistency of our scheme with the underlying probability interpretation of the PF. This is in spite of the fact that rather than probabilities, our probes encode the measurement of \emph{expectation values} only. In particular, they are not derived from probes for outcome probabilities by weighting with outcome values. Consequently, the construction does not come with a specific non-selective probe either, that would otherwise be derivable from the probabilistic probes. Rather, our proposal is to let the \emph{null probe}, i.e., "free evolution" take the role of the non-selective probe. This can be justified in analogy to the practice in the SFQ when the causality axiom is satisfied (which we also assume, see Section~\ref{sec:relcausal}).

The modulus-square construction satisfies a number of desiderata such as positivity, locality and compositionality as already mentioned. It also recovers the results of \emph{single measurements}, which can be expressed already with standard methods of the usual QFT formalism (Section~\ref{sec:single_meas}). Moreover, for quadratic observables we propose a renormalization prescription (Section~\ref{sec:sclassrenorm}). In the special case of a single instantaneous measurement this has the same effect as the well-known normal ordering prescription. In contrast to the latter, however, it generalizes consistently to composite measurements, leading to the slogan "renormalization commutes with composition" (Section~\ref{sec:renormcompose}). Another property that we take as a defining feature of our renormalization prescription is the recovery of semiclassical expectation values for single measurements. That is, we require the expectation value for a single measurement in a coherent state for a given classical solution to be the classical expectation value. This requirement not only can be satisfied, but it uniquely fixes our renormalization scheme.

While most of our treatment applies to free bosonic field theory in globally hyperbolic spacetimes (without, however, taking into account possible gauge symmetries) we specialize to scalar field theory in order to exhibit our construction in first examples in Section~\ref{sec:scalar}. We start with the field observable $\phi^2(x)$ which allows us to show how the subtraction operation in the renormalization scheme can be implemented via point-splitting, as might be expected. More interestingly, we exhibit then the correlation function $\phi^2(x)\phi^2(x')$ as a composite of two measurements of this type (Section~\ref{sec:pointsplit}). Crucially, the word "correlation function", in contrast to its usual use in QFT, does here refer to the actual expectation value of a correlation \emph{in a composite measurement}. In particular, this expectation value is a real quantity as it must be.

The modulus-square construction is sufficiently flexible to also cover more complicated observables involving linear combinations and derivatives. A prime example is the energy-momentum tensor (Section~\ref{sec:emtens}). Building on the (point-split and then renormalized version of the) observable $\phi^2(x)$, we can define the probe measuring the expectation value of the energy-momentum tensor. Note that this inherits the mentioned semiclassicality property. What is more, as a proof of concept, we exhibit the correlation function between the energy momentum tensor evaluated at two different spacetime points. Again, "correlation function" refers here to the expectation value of an actual composite measurement. In particular, this is real by construction.

As a first point of comparison to other approaches we emphasize the versatility of our notion of composition. On the conceptual side this comes from our use of the local PF, while on the implementation side this comes from the path integral. Two measurements can be composed whenever the spacetime regions where they are localized do not overlap.\footnote{Even though the PF does not, the path integral does even permit composition in the case of overlapping regions for the modulus-square construction. However, as the physical meaning of such a composition is unclear, we refrain here from discussing it further.} Note that we have often formally written compositions of measurements in terms of the composition of consecutive time-interval regions in this paper. However, we have done so merely for simplicity of presentation, with the results being valid in the general case. We would also like to emphasize that in the local PF used here there is no need to frame a composition of measurements in terms of a consecutive application of "update rules". The latter may be convenient, and are naturally realized in the SFQ. In general, however, a composition can only be framed in terms of consecutive updates if there are spacelike hypersurfaces separating the regions where the individual measurements are localized. Note that this precludes for example composing the measurements of two detectors that can signal to each other during their lifetimes. A measurement framework that structurally relies on consecutive updates for composition and is thus limited by this separability condition, is that of Fewster and Verch \cite{FeVe:qftlocalmeasure}, based on algebraic quantum field theory.

As mentioned in the introduction, a lot of research has focused on measurement schemes that involve an explicit modeling of the apparatus. It would be highly desirable to provide a connection between observable-based schemes, such as ours, and apparatus-based schemes. Of those approaches where the apparatus is modeled as a quantum field the Fewster-Verch framework is probably the most prominent one. This framework does include a prescription to compute an induced system observable from a given ancilla field (called there "probe field"), interaction and observable on the ancilla. In the other direction, of providing an ancilla field and interaction to measure a specific observable there is also recent progress \cite{MaNa:qftfv}. The latter work could serve as a starting point to establish relations to the approach presented here.

We proceed to consider approaches that involve detector models of the Unruh-deWitt (UDW) type. Our starting point is the observation of a semblance of the square-root construction to equipping the QFT with a modified localized interaction. Inserting a source $J(x)$ into the field is just the same as defining the Weyl observable
\be
F (\phi) = \exp (\im \int\xd^4 x\, J (x) \phi (x))
\label{eq:weylsrc}
\ee
and using the probe $\cP[F|F]$. In the simplest case, UDW detector interactions are just of such a form, where the source is concentrated on the timelike detector trajectory. What is more, the detector has its own degrees of freedom. To take them into account one may make the source into an operator acting on the state space of the detector. However, this would lead outside our present path integral setting. A treatment of the UDW detector fully in terms of a path integral was recently achieved \cite{BPT:pathintUDW}. However, this involved the degrees of freedom of the detector encoded as an additional (fermionic) field in the path integral. In contrast, we are interested in the induced observables in the system, i.e., a path integral purely for the system. Such a treatment was provided in \cite{CoOeZa:udwevanescent}. Fixing the initial state $\psi_\Lini$ and final state $\psi_\Lfin$ of the detector, the typical induced observable takes the form \cite[equation (9)]{CoOeZa:udwevanescent},\footnote{We do not take into account here the renormalization prescription applied in \cite{CoOeZa:udwevanescent}.}
\begin{multline}
    O_{\psi_{\Lini}\to\psi_{\Lfin}}(\phi)=\sum_{n=0}^\infty (-\im)^n \lambda^n \int_{-\infty}^{\infty}\xd \tau_1 \int_{-\infty}^{\tau_1}\xd \tau_2 \cdots \int_{-\infty}^{\tau_{n-1}}\xd \tau_n\\
    \chi(\tau_1)\cdots\chi(\tau_n) f_{\psi_{\Lini}\to\psi_{\Lfin}}(\tau_1,\ldots,\tau_n)\, \phi(x(\tau_1))\cdots\phi(x(\tau_n)) .
    \label{eq:detobs}
\end{multline}
Here, $x(\tau)$ is the detector trajectory in terms of the proper time $\tau$, $\chi(\tau)$ is the switching function, $\lambda$ a coupling constant and the function $f$ depends on initial and final detector states. The corresponding probe is thus $\cP[O_{\psi_{\Lini}\to\psi_{\Lfin}}|O_{\psi_{\Lini}\to\psi_{\Lfin}}]$. This shows that the UDW detector can be plainly treated within the present framework. What is less clear though, is the extent to which the modulus square $|O_{\psi_{\Lini}\to\psi_{\Lfin}}|^2$ can be considered the measured observable. In any case, these could be initial steps towards a more in-depth comparison to a UDW detector based measurement scheme such as the one by Polo-Gómez, Garay and Martín-Martínez \cite{PGGaMM:detectormeasurementqft}.

Another detector based measurement approach was proposed by Anastopoulos, Hu and Savvidou \cite{AHS:qftqi}. In this, probabilities for pointer observables are deduced from the formalism of quantum temporal probabilities. They concretely consider a series of detector apparatuses, each of which interacts with the quantum field. The measurement is modelled as a transition between two complementary subspaces of the Hilbert space, one of which is the subspace of all the states of the apparatus with definite macroscopic records of detection. The measurement induces a transition to this Hilbert subspace and gives a specific pointer observable $\lambda$. This corresponds to a positive operator that interestingly takes the form of a square-root operator $\sqrt{\Pi (\lambda)}$, reminiscent of the form of observables in the modulus-square construction. The probabilities in \cite{AHS:qftqi} are given via $2n$-point correlation functions within the Schwinger-Keldysh closed-time path integral. These correlation functions are "balanced", with each spacetime point appearing twice, once in the forward and once in the backward branch, exactly as those we obtain from composing measurements of field observables $\phi^2(x)$ at different spacetime points, see Section~\ref{sec:pointsplit}. Unsurprisingly, also the generating functional used \cite[equation~(35)]{AHS:qftqi}, takes the form of the probe $\cP[F|F]$ for a Weyl observable $F$ determined by sources similar to \eqref{eq:weylsrc}. This striking convergence with our approach suggests a deeper relation.

The presented approach has a number of limitations. First of all the probes we construct, exclusively encode measurements of expectation values (and their multiplicative composites), rather than probabilities. Moreover, the latter cannot be deduced in a simple manner from the former. The situation is akin to measuring expectation values of observables in SQT when only the corresponding operator is available, but not its spectral decomposition. This suggests applying the same remedy applicable there, namely constructing probes from the spectral decomposition. In an operator setting this was indeed done recently for field operators \cite{Oe:spectral}. This suggests transferring those results to our present setting of the local PF using the path integral prescription. In this context it turns out that the probe that measures whether a linear observable $D$ has the value $q$ on the space of field configurations, can be approximately expressed precisely by the modulus-square construction for the following observable:
\begin{equation}
    H_{D}^{\epsilon}(q)(\phi)
    \defeq\frac{1}{\sqrt{\pi}\epsilon}\exp\left(-\frac{1}{\epsilon^2}(D(\phi)-q)^2\right) .
    \label{eq:obsdq}
\end{equation}
That is, the corresponding probe is given by, $\cP[H_{D}^{\epsilon}(q)|H_{D}^{\epsilon}(q)]$. The exact correspondence with the spectral decomposition is obtained in the limit $\epsilon\to 0$. By integrating over $q$ with suitable functions inserted we can derive probes for related measurements,
\begin{equation}
    \sqrt{2\pi}\epsilon\int_{-\infty}^{\infty}\xd q\, f(q) \cP[H_{D}^{\epsilon}(q)|H_{D}^{\epsilon}(q)] .
    \label{eq:probesd}
\end{equation}
The expectation value would be obtained by $f(q)=q$, while characteristic functions for subsets of $\R$ would yield probabilities for the outcome to lie in the respective subset. Crucially, we also obtain a proper non-selective probe by setting $f(q)=1$. Note also that within the spirit of our present modulus-square construction the classical observable computed in the case of the expectation value is recovered correctly,
\begin{equation}
    \sqrt{2\pi}\epsilon\int_{-\infty}^{\infty}\xd q\, q |H_{D}^{\epsilon}(q)(\phi)|^2=D(\phi) .
\end{equation}
(This fixes the prefactor in \eqref{eq:probesd}.)
Clearly, the observable \eqref{eq:obsdq} provides an example of the modulus-square construction very different from those envisaged in the body of this work. This shows the potential to alleviate another limitation of the presented approach, namely the restriction to a special class of observables.

\subsection*{Acknowledgments}

This work was partially supported by UNAM-PAPIIT project grant IN106422 and UNAM-PASPA-DGAPA. A.Z. acknowledges support by the UNAM Postdoctoral Program (POSDOC) and by the National Science Centre in Poland under the research grant Maestro (2021/\!\! 42/\!\! A/\!\! ST2/\!\! 00356).

\subsection*{Author contributions}

Both authors contributed equally to the present work.

\appendix


\section{Observable propagators}
\label{sec:vacobs}

In the present section we establish basic definitions and relations of \emph{observable propagators}. These are generalizations of the usual \emph{field propagators} where observables $\phi\mapsto\phi(x)$ (e.g., in Klein-Gordon theory) are replaced by general linear observables (in arbitrary linear bosonic field theories). In particular, we are interested in the vacuum correlation function of a product of two linear observables $D,D'$,
\begin{equation}
  \wuu(D,D')\defeq \rho[D D'](\ncoh_0)=\int\xD\phi\, D(\phi) D'(\phi) e^{\im S(\phi)} .
  \label{eq:obsfprop}
\end{equation}
Note that this bilinear form is symmetric by definition, $\wuu(D,D')=\wuu(D',D)$.
For point observables $D_x(\phi)=\phi(x)$ and $D_{x'}(\phi)=\phi(x')$ in Klein-Gordon theory this is the usual time-ordered (Feynman) propagator,
\begin{equation}
  \wtuu(x,x')\defeq \wuu(D_{x},D_{x'})= \rho[D_x D_{x'}](\ncoh_0)
  = \langle\vac,\mathbf{T} \phi(x') \phi(x) \vac\rangle
  =  -\im G_F(x,x').
 \label{eq:time-orderedprop}
\end{equation}
Similarly, we call $\wuu$ the \emph{time-ordered observable propagator}.

We use methods and conventions of \cite{Oe:feynobs} and \cite{CoOe:locgenvac}. 
First recall that the complex structure on phase space also induces a \emph{polarization} which is a direct sum decomposition of the complexified phase space into conjugate subspaces $L^{\C}=L ^+ \oplus L^-$. Usually, $L^-$ and $L^+$ are called respectively the spaces of positive and negative energy solutions. We write $\phi=\phi^+ + \phi^-$ for the decomposition of elements of $L$ into positively and negatively polarized components. We recall the identity,
\begin{equation}
\{\phi',\phi\}=4\im\omega(\phi'^-,\phi^+) .
\end{equation}
Consider now a spacetime region determined by a time interval $[t_1,t_2]$ and a linear observable $D:K_{[t_1,t_2]}\to\R$ located within. If we modify the action by adding the observable $D$ to it, the resulting homogeneous equations of motions are replaced by inhomogeneous ones. We denote by $\eta$ the unique complexified solution to these inhomogeneous equations of motion that satisfies the Feynman boundary conditions, i.e., that is negatively polarized in the future of the support of $D$ and positively polarized in its past. Note that $\eta$ is homogeneous outside of the support of $D$. We denote by $\eta_1$ and $\eta_2$ the induced germs of solutions on the equal-time hypersurfaces at times $t_1$ and $t_2$ respectively. As usual, we identify the spaces of germs of solutions for any Cauchy hypersurface with the phase space $L$ of global solutions. In particular, $\eta_1\in L^+$ and $\eta_2\in L^-$. Define $\tau\defeq \eta_2-\eta_1$. Then, $\tau^+=-\eta_1$ and $\tau^{-}=\eta_2$. If we evaluate the observable $D$ on a solution $\phi$ that is homogeneous in $[t_1,t_2]$ we obtain,
\begin{equation}
  D(\phi)=2\omega_{\partial [t_1,t_2]}(\phi,\eta)
  =2\omega_{t_1}(\phi_1,\eta_1)-2\omega_{t_2}(\phi_2,\eta_2)
  =2\omega(\phi,\eta_1-\eta_2)=2\omega(\tau,\phi) .
  \label{eq:dxi}
\end{equation}
Here $\omega_{\partial [t_1,t_2]}$ denotes the symplectic form on the boundary of the time interval region determined by $[t_1,t_2]$, which decomposes into the initial hypersurface at $t_1$ and the final hypersurface at $t_2$. The minus sign in the second equality arises from the relative change of orientation of the final hypersurface compared to the initial one as boundary components of $[t_1,t_2]$. Further, due to homogeneity we have $\phi_1=\phi_2=\phi$, using a simplified notation, where global solutions and their germs on Cauchy hypersurfaces are denoted by the same symbol. Note that since $D$ is real by assumption, $\tau$ is also real by relation \eqref{eq:dxi}, i.e., it is an element of the ordinary phase space $L$ and not only of its complexification $L^{\C}$.

The observable propagator \eqref{eq:obsfprop} can be expressed in terms of the inhomogeneous solutions as follows \cite{Oe:feynobs,CoOe:locgenvac},\footnote{The expression in the references corresponds to $\wuu(D,D)=-\im D(\eta)$. The stated expression follows with bilinearity and symmetry.}
\begin{equation}
  \wuu(D,D')=-\frac{\im}{2} (D(\eta')+D'(\eta)) .
\end{equation}
To calculate this we first suppose that the second observable $D'$ has support later than $D$. Thus, let $D':K_{[t_2,t_3]}\to\R$ with support in the spacetime region given by the time interval $[t_2,t_3]$. Let $\eta'$ be the corresponding inhomogeneous solution in spacetime. We set $\tau'\defeq\eta'_3-\eta'_2$ etc. Noting $\eta'_1=\eta'_2$, the evaluation of $D$ on $\eta$ yields,
\begin{equation}
  D(\eta')=2\omega_{\partial [t_1,t_2]}(\eta',\eta)
  =2\omega_{t_1}(\eta'_1,\eta_1)-2\omega_{t_2}(\eta'_2,\eta_2)
  =2\omega(\tau,\eta'_2)=-2\omega(\tau,\tau'^+)=\frac{\im}{2}\{\tau,\tau'\} .
\end{equation}
Instead, evaluating $D'$ on $\eta'$ yields the same result (noting $\eta_2=\eta_3$),
\begin{equation}
    D'(\eta)=2\omega_{\partial [t_2,t_3]}(\eta,\eta')
    =2\omega_{t_2}(\eta_2,\eta'_2)-2\omega_{t_3}(\eta_3,\eta'_3)
    =2\omega(\tau',\eta_2)=2\omega(\tau',\tau^-)=\frac{\im}{2}\{\tau,\tau'\} .
\end{equation}
This motivates the definition of the following observable propagators:
\begin{equation}
  \wud(D,D')\defeq\wdu(D',D)\defeq\frac12\{\tau,\tau'\} .
  \label{eq:updownprop}
\end{equation}
We extend this definition to the case where $D$ and $D'$ are arbitrarily localized relative to each other. Note
\be
\wud(D,D')=\overline{\wud(D',D)}.
\label{eq:wightmanconj}
\ee
The relation to $\wuu$ is then,
\begin{itemize}
  \item
  If $D\ll D'$ ($D$ earlier than $D'$), then $\wuu(D,D')=\wud(D,D')=\wdu(D',D)$.
  \item
  If $D\gg D'$ ($D$ later than $D'$), then $\wuu(D,D')=\wud(D',D)=\wdu(D,D')$.
\end{itemize}
In the case of scalar point observables we recover the usual \emph{Wightman propagator},
\begin{equation}
  \wtud(x,x')\defeq \wud(D_{x},D_{x'})=\langle\vac,\phi(x') \phi(x) \vac\rangle .
\end{equation}

It remains to consider $\wuu$ in the case that neither $D$ is later than $D'$ nor the other way round. To this end we suppose that we can approximate these observables by linear combinations of instantaneous observables,
\begin{equation}
  D=\sum_{k=1}^n D_k,\qquad D'=\sum_{j=1}^m D'_j .
\end{equation}
We have then,
\begin{align}
  \wuu(D,D')& =\sum_{k=1}^n \sum_{j=1}^m \begin{cases}
     \frac12 \{\tau_k,\tau'_j\} & \text{if}\; D_k\ll D'_j\\
     \frac12 \{\tau'_j,\tau_k\} & \text{if}\; D_l\gg D'_j
  \end{cases} \\
  & =\frac12 g(\tau,\tau') + \im \sum_{k=1}^n \sum_{j=1}^m \begin{cases}
    \omega(\tau_k,\tau'_j) & \text{if}\; D_k\ll D'_j\\
    \omega(\tau'_j,\tau_k) & \text{if}\; D_k\gg D'_j
 \end{cases}
\end{align}
Crucially, the real part of the propagator $\wuu(D,D')$ does not depend on the temporal order of the observables and their components, while the imaginary part changes sign under a change of order. The latter property motivates the definition of the anti-time-ordered propagator,
\begin{equation}
  \wdd(D,D')\defeq \overline{\wuu(D,D')} .
  \label{eq:time-antitime_relation}
\end{equation}
Thus, we have in all cases that a complex conjugation corresponds to the reversal of all arrow directions in the notation. Moreover, an exchange of the two arguments corresponds to an exchange of the two arrows.
We also denote the real part by,
\begin{equation}
  r(D,D')\defeq \Re(\wuu(D,D'))=\Re(\wdd(D,D')) = \Re(\wud(D,D'))=\Re(\wdu(D,D')) .
\end{equation}
We also have the identity,
\begin{equation}
  r(D,D')=\frac12\left(\wud(D,D')+\wud(D',D)\right) .
\end{equation}
In the case of scalar point observables we can write this as the symmetrized Wightman propagator,
\begin{equation}
  \rp(x,x')\defeq \frac12\left(\langle\vac,\phi(x') \phi(x) \vac\rangle +\langle\vac,\phi(x) \phi(x') \vac\rangle\right) .
\end{equation}


\section{Modulus-square probes with discard}
\label{sec:derivations}

\subsection{Weyl observables}

In this section, we explicitly calculate the composition of the modulus-square probe for Weyl observables in a time-interval, compare expression \eqref{eq:mspfact}, with the discard. This serves as a generating function for modulus-square probes for polynomial observables composed with the discard.  Again we use methods of \cite{Oe:feynobs} and \cite{CoOe:locgenvac}.
First, we rewrite \eqref{eq:mspfact} in a form that makes manifest that the region $M$ is a time-interval $[t_1,t_2]$ by splitting boundary coherent states into initial and final coherent states,
\begin{multline}
    \cP_{[t_1,t_2]}[F | F'](\mcoh_{\phi|\phi'}\tens\mcoh_{\eta|\eta'}) \\
    =\np_{[t_1,t_2]}(\mcoh_{\phi|\phi'}\tens\mcoh_{\eta|\eta'}) F((\phi,\eta)^{\Lint}) \overline{F'((\phi',\eta')^{\Lint})} 
    \exp\left(-\frac12 \wuu(D,D) -\frac12 \overline{\wuu(D',D')}\right) .
    \label{eq:mspfactt}
\end{multline}
As we have seen previously, we have $(\phi,\eta)^{\Lint} = \phi^- +\eta^+$. This allows us to factorize,
\begin{equation}
    F((\phi,\eta)^{\Lint}) \overline{F'((\phi',\eta')^{\Lint})}
    = F(\phi^-)\overline{F'(\phi'^-)} F(\eta^+)\overline{F'(\eta'^+)} .
\end{equation}
Associating to $D,D'$ the corresponding elements $\tau,\tau'\in L$ as in Appendix~\ref{sec:vacobs} and using relation \eqref{eq:dxi}, we can rewrite,
\begin{multline}
    F(\eta^+)\overline{F'(\eta'^+)}=\exp\left(\im D(\eta^+)-\im D'(\eta'^-)\right) \\
    =\exp\left(2\im\omega(\tau,\eta^+)-2\im\omega(\tau',\eta'^-)\right)
    =\exp\left(\frac12\{\tau,\eta\}+\frac12\{\eta',\tau'\}\right) .
\end{multline}
Evaluating the null probe using \eqref{eq:evoltonp} and noting that the coherent states are parametrized in a time-independent way yields,
\begin{multline}
    \np_{[t_1,t_2]}(\mcoh_{\phi|\phi'}\tens\mcoh_{\eta|\eta'})
    =\langle \ncoh_{\eta},\ncoh_{\phi}\rangle \overline{\langle \ncoh_{\eta'},\ncoh_{\phi'}\rangle} \\
    =\exp\left(\frac12\{\phi,\eta\}+\frac12\{\eta',\phi'\}
    -\frac14\{\phi,\phi\}-\frac14\{\phi',\phi'\}-\frac14\{\eta,\eta\}-\frac14\{\eta',\eta'\}\right) .
\end{multline}

Next, we realize that applying the discard, i.e.\ the trace, on the final state is, due to the completeness relation of coherent states \eqref{eq:completeness} equivalent to an integral, as follows,
\begin{equation}
    \discard\comp\cP_{[t_1,t_2]}[F | F'](\mcoh_{\phi|\phi'})
    =\int_{\hat{L}}\xd\nu(\eta) \exp\left(\frac12\{\eta,\eta\}\right)
     \cP_{[t_1,t_2]}[F | F'](\mcoh_{\phi|\phi'}\tens\mcoh_{\eta|\eta}) .
\end{equation}
We may now evaluate this integral using expression \eqref{eq:mspfactt}. We focus only on the part of the integrand dependent on $\eta$,
\begin{align}
    & \int_{\hat{L}}\xd\nu(\eta) \exp\left(\frac12\{\eta,\eta\}\right)
    \np_{[t_1,t_2]}(\mcoh_{\phi|\phi'}\tens\mcoh_{\eta|\eta})
    F(\eta^+)\overline{F'(\eta^+)} \\
    & =\int_{\hat{L}}\xd\nu(\eta)
    \exp\left(\frac12\{\tau+\phi,\eta\}+\frac12\{\eta,\tau'+\phi'\}
    -\frac14\{\phi,\phi\}-\frac14\{\phi',\phi'\}  \right) \\
    & = \exp\left(\frac12\{\tau+\phi,\tau'+\phi'\}
    -\frac14\{\phi,\phi\}-\frac14\{\phi',\phi'\}  \right) \\
    & = \exp\left(\frac12\{\phi,\phi'\}
    -\frac14\{\phi,\phi\}-\frac14\{\phi',\phi'\}
    +\frac12\{\tau,\phi'\}+\frac12\{\phi,\tau'\}+\frac12\{\tau,\tau'\}  \right) \\
    & = \langle \ncoh_{\phi'},\ncoh_{\phi}\rangle F(\phi'^+) \overline{F'(\phi^+)}
    \exp\left(\wud(D,D')\right) \\
    & =  \discard\comp\np_{[t_1,t_2]}(\mcoh_{\phi|\phi'}) F(\phi'^+) \overline{F'(\phi^+)}
    \exp\left(\wud(D,D')\right) .
\end{align}
With this we obtain in total,
\begin{multline}
    \discard\comp\cP_{[t_1,t_2]}[F | F'](\mcoh_{\phi|\phi'}) 
    =\discard\comp\np_{[t_1,t_2]}(\mcoh_{\phi|\phi'}) F(\phi^- +\phi'^+) \overline{F'(\phi^+ +\phi'^-)} \\
    \exp\left(\wud(D,D')-\frac12 \wuu(D,D) -\frac12 \wdd(D',D')\right) .
    \label{eq:weylprobe}
\end{multline}
We can rewrite the observable evaluation part as,
\begin{equation}
    F(\phi^- +\phi'^+) \overline{F'(\phi^+ +\phi'^-)}
    =\exp\left(\im (D-D')(\phi^- +\phi'^+)\right) .
\end{equation}
Finally, we note that if we set $D=D'$ (and thus $F=F'$) the results simplifies considerably as the probe $\cP_{[t_1,t_2]}[F | F]$ becomes non-selective,
\begin{equation}
\discard\comp\cP_{[t_1,t_2]}[F | F](\mcoh_{\phi|\phi'})=\discard\comp\np_{[t_1,t_2]}(\mcoh_{\phi|\phi'})=\tr(\mcoh_{\phi|\phi'})=\langle \ncoh_{\phi'},\ncoh_{\phi}\rangle .
\end{equation}
Of course, for the use as a generating function, it is crucial to allow $D$ and $D'$ to be different.

\subsection{Some composite observables}
\label{sec:quadobs}

In this section we calculate some composite modulus-square probes with discard.  We use formula \eqref{eq:weylprobe} combined with the generating function method \eqref{eq:functional_deriv_weyl}.
For a single quadratic observable $\Lambda=D^2$, but allowing the two branches to be different, $D\neq D'$, we obtain, 
\begin{multline}
  \discard\comp\cP_{[t_1,t_2]}[D | D'](\mcoh_{\phi|\phi'})
  = \frac{\partial}{\partial\lambda} \frac{\partial}{\partial\lambda'}
  \discard\comp\cP_{[t_1,t_2]}[\exp(\im\lambda D)| \exp(\im\lambda' D')](\mcoh_{\phi|\phi'}) \big|_{\lambda,\lambda'=0} \\
  =\discard\comp\np_{[t_1,t_2]}(\mcoh_{\phi|\phi'})
   \left(D(\phi^- + \phi'^+) D'(\phi^- + \phi'^+) + \wud(D,D')\right).
   \label{eq:probe_quadr_bare}
\end{multline}
With $D'=D$ and $\phi'=\phi$ we recover expression \eqref{eq:probe_discard}.

We proceed to consider the product of two quadratic observables $\Lambda_1=D_1^2$ and $\Lambda_2=D_2^2$. Again we allow the linear observables to be different in the two branches.
\begin{multline}
  \discard\comp\cP_{[t_1,t_2]}[D_1 D_2 | D'_1 D'_2](\mcoh_{\phi|\phi'}) \\
  = \frac{\partial}{\partial\lambda_1} \frac{\partial}{\partial\lambda_1'}
    \frac{\partial}{\partial\lambda_2} \frac{\partial}{\partial\lambda_2'}
  \discard\comp\cP_{[t_1,t_2]}[\exp\left(\im(\lambda_1 D_1 +\lambda_2 D_2)\right)| \exp\left(\im(\lambda_1' D_1'+\lambda_2' D_2')\right)](\mcoh_{\phi|\phi'}) \big|_{\lambda_1,\lambda_1',\lambda_2,\lambda_2'=0} \\
  =\discard\comp\np_{[t_1,t_2]}(\mcoh_{\phi|\phi'})
  \Big(
    \wud (D_1,D_2') \wud (D_2,D_1')
    + \wud (D_1 ,D_1')  \wud (D_2 ,D_2') \\
    + \wuu(D_1, D_2)
    \wdd(D_1', D_2')
    +\wud (D_1,D_1') 
    \wD_2 
    \wD_2' 
    + \wud (D_2,D_2')
    \wD_1 
    \wD_1'
    + \wud (D_1,D_2')
    \wD_2 
    \wD_1' \\
    +\wud (D_2 ,D_1')
    \wD_1
    \wD_2'
    + \wuu(D_1, D_2)
    \wD_1' 
    \wD_2' 
    + \wdd(D_1', D_2')
    \wD_1 
    \wD_2 
    +\wD_1
    \wD_2
    \wD_1'
    \wD_2' 
    \Big) .
    \label{eq:dprobe}
\end{multline}
To simplify the notation we have used abbreviations of the form $\widetilde{D}=D(\phi^{-}+\phi'^{+})$.

Finally, we consider the composition of a Weyl observable $F_1=\exp(\im D_1)$ with a single quadratic observable $D_2$,
\begin{multline}
  \discard\comp\cP_{[t_1,t_2]}[F_1 D_2 | F'_1 D'_2](\mcoh_{\phi|\phi'}) \\
  = \frac{\partial}{\partial\lambda_2} \frac{\partial}{\partial\lambda_2'}
  \discard\comp\cP_{[t_1,t_2]}[\exp\left(\im( D_1 +\lambda_2 D_2)\right)| \exp\left(\im( D_1'+\lambda_2' D_2')\right)](\mcoh_{\phi|\phi'}) \big|_{\lambda_2=\lambda_2'=0} \\
  =\discard\comp\np_{[t_1,t_2]}(\mcoh_{\phi|\phi'})
  \exp (\im \wD_1 -\im \wD_1' + \wud (D_1,D_1') -\frac{1}{2} (\wuu (D_1,D_1) + \wdd (D_1',D_1'))) \\
\left(\left(\im \wD_2 +\wud(D_2,D_1')-\wuu(D_1,D_2)\right)
\left(-\im\wD'_2 +\wud(D_1,D_2') -\wdd(D_1',D_2')\right)
+\wud (D_2,D_2')\right) .
\end{multline}
If $D_1=D_1'$, this simplifies considerably,
\begin{multline}
  \discard\comp\cP_{[t_1,t_2]}[F_1 D_2 | F_1 D'_2](\mcoh_{\phi|\phi'})
  =\discard\comp\np_{[t_1,t_2]}(\mcoh_{\phi|\phi'}) \\
\left(\left(\im \wD_2 +\wud(D_2,D_1)-\wuu(D_1,D_2)\right)
\left(-\im\wD'_2 +\wud(D_1,D_2') -\wdd(D_1,D_2')\right)
+\wud (D_2,D_2')\right) .
\label{eq:pid3}
\end{multline}

\bibliographystyle{quantumc}
\bibliography{stdrefsb}
\end{document}